\documentclass{aa}  

\usepackage{graphicx}
\usepackage{xcolor}
\usepackage{txfonts}
\usepackage{multirow}
\usepackage{rotating}

\usepackage[hidelinks,colorlinks=true,linkcolor=red,citecolor=blue,urlcolor=violet]{hyperref}
\usepackage{natbib}

\newcommand{\cn}[1]{\textbf{{\color{magenta}{{[xxx CN: #1]}}}}}

\newcommand{\upeak}{u_{\rm{peak}}}
\newcommand{\nee}{n_{\rm{e}}}
\newcommand{\Tsl}{T_{\rm{sl}}}
\newcommand{\Tgas}{T_{\rm{gas}}}
\newcommand{\nH}{n_{\rm{H}}}
\newcommand{\Dang}{D_{\rm{ang}}}

\newcommand{\Prob}{\mathcal P}

\newcommand{\Rhat}{\hat R}
\newcommand{\phihat}{\hat \phi}
\newcommand{\Ri}{\hat R_{\mathrm{i}}}
\newcommand{\Rplus}{\hat R_{\mathrm{i + 1}}}
\newcommand{\radj}{r_{\mathrm{j}}}
\newcommand{\radplus}{r_{\mathrm{j + 1}}}
\newcommand{\Nbins}{N_{\mathrm{bins,X}}}
\newcommand{\NSZ}{N_{\mathrm{bins,SZ}}}

\newcommand{\gamIN}{\gamma'_{\mathrm{IN}}}
\newcommand{\gamOUT}{\gamma'_{\mathrm{OUT}}}

\newcommand{\Rbreak}{R_{\mathrm{break}}}
\newcommand{\Mtwo}{M_{200}}
\newcommand{\Mtwobar}{{M}_{200}^{\mathrm{median}}}
\newcommand{\rtwobar}{{r}_{200}^{\mathrm{median}}}
\newcommand{\rtwo}{r_{200}}
\newcommand{\Mfive}{M_{500}}
\newcommand{\rfive}{r_{500}}
\newcommand{\ctwo}{\ln c_{200}}
\newcommand{\fff}{\epsilon_{\rm{N}}}
\newcommand{\gggg}{\epsilon_{\rm{T}}}
\newcommand{\sss}{\epsilon_{\rm{SZ}}}

\newcommand{\Msol}{\mathrm M_{\mathrm{\odot}}}

\newcommand{\sigmastat}{\sigma_{\mathrm{stat}}} 

\newcommand{\ptot}{p}
\newcommand{\ptotstar}{p_{\star}}
\newcommand{\alphaturb}{\alpha_{\rm{turb}}}
\newcommand{\pturb}{p_{\rm{turb}}}
\newcommand{\alphainf}{\alpha_{\rm{inf}}}
\newcommand{\alphazero}{\alpha_0}

\newcommand{\rhocrit}{\rho_{\rm{crit}}}

\newcommand{\kboltz}{k_{\rm B}}
\newcommand{\proton}{m_{\rm p}}
\newcommand{\Phieff}{\Phi_{\rm{eff}}}
\newcommand{\Tsp}{T_{\rm{sp}}}
\newcommand{\Tsli}{T_{\rm{sl,i}}}

\newcommand{\Normi}{Norm_{\rm{i}}}

\newcommand{\nstar}{n_\star}

\newcommand{\PSZ}{P_{\rm{SZ}}}
\newcommand{\PSZj}{P_{\rm{SZ,j}}}

\newcommand{\NNN}{\mathcal N}
\newcommand{\Niii}{\mathcal N_{\rm{i}}}
\newcommand{\arcminute}{\rm{arcmin}}
\newcommand{\rout}{r_{\rm{NSZ}}}

\newcommand{\sigmagas}{\sigma_{\rm{gas,1D}}}
\newcommand{\sigmaturb}{\sigma_{\rm{turb,1D}}}

\newcommand{\kms}{\rm{km \, s^{-1}}}

\newcommand{\niii}{n_{\rm{i}}}
\newcommand{\Rhatend}{\hat R_{\rm{NX}}}
\newcommand{\diff}{\mathrm{d}}

\newcommand{\alphargs}{\alpha_{\rm{0,RGS}}}
\newcommand{\sigmargs}{\sigma_{\rm{RGS}}}
\newcommand{\Rpeak}{R_{\rm{peak}}}

\begin{document} 

\defcitealias{Bartalesi24}{B24}
\defcitealias{Ghirardini19}{G19}
\defcitealias{angelinelli20}{A20}
\defcitealias{Ettori22}{E22}

   \title{Searching for rotation in X-COP galaxy clusters}

   \author{T. Bartalesi\inst{1,2}\fnmsep\thanks{\email{tommaso.bartalesi3@unibo.it}}
          \and
          S. Ettori\inst{2,3}
          \and
          C. Nipoti\inst{1}
          }


   \institute{Dipartimento di Fisica e Astronomia “Augusto Righi” – Alma Mater Studiorum – Università di Bologna, via Gobetti 93/2, I-40129 Bologna 
         \and INAF, Osservatorio di Astrofisica e Scienza dello Spazio, via Piero Gobetti 93/3, 40129 Bologna, Italy 
         \and INFN, Sezione di Bologna, viale Berti Pichat 6/2, 40127 Bologna, Italy
             }

\date{Accepted March 13, 2025}

  \abstract
   {}
   {
We search for evidence of rotational support by analyzing the thermodynamic profiles of the intracluster medium (ICM) in a sample of nearby, massive galaxy clusters.
   }
   {
   For each object of the {\it XMM-Newton} Cluster Outskirts Project (X-COP) sample, we present axisymmetric models of rotating ICM with composite polytropic distributions, in equilibrium in spherically symmetric dark halos, exploring cases both with and without turbulent support in the ICM. The profile of rotation velocity and the distribution of turbulent velocity dispersion are described with flexible functional forms, consistent with the properties of synthetic clusters formed in cosmological simulations.
   The models are tuned via a Markov Chain Monte Carlo algorithm to reproduce the radial profiles of the thermodynamic variables as resolved in the \textit{XMM-Newton} and \textit{Planck} maps, and to be consistent with the mass distributions estimated either from weak lensing observations (when available) or under the assumption of a "universal" value of the baryon fraction.
   }
    {Our models indicate that there is room for non-negligible rotation in the ICM of massive clusters, with typical peak rotation speed $\approx$ 300 $\kms$ and peak rotation-velocity-to-velocity-dispersion ratio $u_\phi/\sigmagas\approx 0.3$. 
    According to our models, the ICM in Abell~2255 can have a rotation speed as high as 500 $\kms$, corresponding to $u_\phi/\sigmagas\approx 0.4$, at a distance of 100 kpc from the center, where the X-ray emissivity is still high. 
    This makes Abell~2255 a very promising candidate for the presence of rotation in the ICM that could be detected with the currently operating XRISM observatory, as we demonstrate computing and analyzing a mock X-ray spectrum.
    }
{}
   \keywords{Galaxies: clusters: general -- Galaxies: clusters: individual: Abell~2255 --- Galaxies: clusters: intracluster medium -- X-rays: galaxies: clusters}

   \maketitle
%

\section{Introduction}
\label{sec:intro}

Clusters of galaxies are permeated by hot ($\sim 10^7-10^8$ K), rarefied ($\sim 10^{-2}-10^{-4}$ ions per cm$^3$), optically thin gas, known as the intracluster medium (ICM), which is believed to be approximately in equilibrium in the total gravitational potential of the system, dominated by the dark-matter halo. The study of the ICM is key to measuring the parameters that define the cosmological background \citep[e.g.][]{Pratt19} and to understanding tnon-gravitational physical processes, such as the mechanisms that prevent the gas from cooling and forming stars in the cluster core. A major limitation to the determination of the cosmological parameters and to the understanding of the ongoing heating mechanisms in the ICM is the poor possibility of gathering information on the velocity field (that is, turbulent and bulk motions) of the gas.

The ICM emits in the X-rays via thermal Bremsstrahlung and the emission lines of the heavy metals \citep[e.g.][for a review]{Sanders23}, and distorts the cosmic microwave background (CMB) through the inverse Compton scattering of ICM electrons, known as the Sunyaev-Zeldovich \citep[SZ;][]{Sunyaev72} effect \citep[see][for a review]{mroczkowski19}.
Current understanding of the ICM thus comes from the spectral analyses of the broad-band X-ray thermal continuum and the assessment of the CMB main distortions.
In recent years, several observational campaigns have aimed to advance our understanding of the thermodynamic properties of the ICM from the center out to the outskirts of clusters. 
These efforts have been focused on nearby, massive clusters, leveraging the joint power of X-ray and SZ effect data from the \textit{XMM-Newton} and \textit{Planck} observatories, respectively.

The limited sensitivity of microwave detectors to the weak distortions in the CMB spectrum caused by the bulk and turbulent motions of electrons in the ICM, known as the kinetic SZ effect \citep[see][]{Baldi18,mroczkowski19, Altamura23} and the degeneracy of this signal with the intrinsic properties of the ICM (like the gas density and geometry) did not allow the characterization of the velocity field in a large sample of objects.
Though the poor spectral resolution of X-ray CCD-like spectrometers (such as \textit{XMM-Newton}) has hindered the detection of the broadening and shifting of the X-ray emitting lines, several attempts have limited the motions of the ICM to few 100 $\kms$ in the inner regions, except for few merging systems with velocities above 500 $\kms$ \citep[e.g.][]{Tamura14, Liu19, sanders20,Gatuzz22b,Gatuzz22,Gatuzz24}.
Overall, the limitations of the available instruments have prevented a comprehensive knowledge of the bulk and turbulent motions in the ICM from being achieved, from the inner region, where, according to numerical simulations, galaxy-scale processes routinely inject significant kinetic energy into the ICM, out to the virialization region, where substantial kinetic energy is expected to be injected by large-scale structure processes \citep[e.g.][]{Vazza12, angelinelli20}.

Precious information on non-thermal motions in the ICM came from the spectra obtained by X-ray Multi-Mirror reflecting grating spectrometer (RGS) onboard \textit{XMM-Newton} in the 0.3--2.5 keV energy range.
However, due to the slitless nature of the detector, the extent of the source contributes to the line width. By assuming that the objects under exam (typically the cores of galaxy clusters) are point sources, some upper limits of $\approx 500\, \kms$ on the velocity broadening are obtained \citep[][for a review]{Sanders11, Pinto15, Bambic18, Sanders23}.

With the advent of the X-ray spectrometer \textit{Resolve} at a high spectral resolution onboard the X-ray Imaging and Spectroscopy Mission \citep[\textit{XRISM};][]{Tashiro18}, measurements of the bulk and turbulent motions in the ICM are now possible \citep[see, e.g.,][hereafter \citetalias{Bartalesi24}]{OTA18, Bartalesi24}. 
However, due to the relatively low collective area and small Field of View of \textit{XRISM}, they are limited to the cores of nearby, massive clusters, where the signal is higher.
The published measurements of the turbulent velocity dispersion of $164 \pm 10\, \kms$ \citep[obtained with \textit{Hitomi}, the predecessor of \textit{XRISM},][]{Hitomi16} and $169 \pm 10\, \kms$ \citep{XRISM_A2029_1} in the cores of the massive clusters Perseus and A2029, respectively, placed an upper limit of 0.04 on the non-thermal-to-total energy ratio in the central regions of these objects, which is an indication of dynamically unimportant turbulence in the inner regions of massive galaxy clusters.

Current knowledge on the mass distribution of galaxy clusters primarily comes from gravitational lensing, and, in the outskirts, from statistical measurements of the weak lensing (WL) signal, that is the mild distortion of the shape of background galaxies  \citep[for a review see][]{Umetsu20}.
However, WL mass measurements for nearby massive objects can be complicated due to their large angular size in the sky. 
Given that the median value of the baryon fraction and the scatter around it agree far beyond the core between samples of clusters formed in cosmological hydrodynamical simulations with different solvers and subgrid models, an independent, but indirect estimate of the cluster mass can be obtained assuming a ``universal'' value of the gas mass fraction as inferred from these simulations \citep{Ghirardini18,Eckert19,Ettori22}.   
Since early measurements of the mass of galaxy clusters, there has been solid evidence that the masses obtained under the assumption of the hydrostatic equilibrium are biased low compared to the WL masses \citep[e.g.][]{Pratt19, Ettori19}. 
This mass discrepancy, the so-called hydrostatic mass bias, has been attributed to the inability to account for some unresolved, residual kinetic energy in the ICM. 
This discrepancy can potentially explain much of the current tension in determining cosmological parameters: between those obtained from cluster halo abundance calibrated on hydrostatic masses and those inferred from modeling the primary anisotropies of the cosmic microwave background \citep{P14}.


The results of hydrodynamic cosmological simulations indicate that both rotation and turbulence are expected to contribute to the velocity field of the ICM \citep[e.g.][]{lau09, Suto13, BALDI17, Braspenning25}. However, there is no general consensus on how the non-thermal support is expected to be split between rotation and turbulence, because of the dependence on the specific implementation of some baryonic physics processes in the hydrodynamic codes. For instance, as far as rotation is concerned, this uncertainty in the predictions of cosmological simulation is illustrated by figure~3 of \citetalias{Bartalesi24}, where the average ICM rotation velocity profiles obtained in two different cosmological simulations are shown (see also \citealt{Suto13,Braspenning25}).

The possibility that the ICM rotates significantly is interesting also for reasons that go beyond gauging the non-thermal support in clusters.  In particular, given that the ICM is weakly magnetized \citep[e.g.][]{Bruggen13}, the rotation of the ICM could be relevant to the energy balance of the gas in galaxy clusters, because the magnetorotational instability \citep{Balbus1991} could be at work in a magnetized rotating ICM \citep{Nipoti2014,Nipoti15}. The non-linear evolution of the magnetorotational instability is expected to lead to turbulent heating, which could contribute to regulating the thermal evolution of the ICM, cooperating with feedback from active galactic nuclei (AGN) in offsetting the radiative cooling of the ICM in the central regions of cool-core clusters \citep[e.g.][]{McNamara12,Hlavacek22}.

In the attempt to improve our understanding of the velocity field of the ICM, in this paper we present cluster models that allow for the presence of rotation and turbulence in the ICM.
Using a Bayesian model-data comparison,  these models are tuned to reproduce the radial profiles of the thermodynamic properties of the ICM as resolved in the XMM-\textit{Newton} and \textit{Planck} maps for the clusters of the XMM-Newton Cluster Outskirts Project (X-COP) sample \citep{X-COP}, and to be consistent with the available virial mass estimates. 
Among the studied clusters, A2255 turns out to be the most promising object in which ICM rotation could be detectable with currently available X-ray spectrometers: we present the mock spectrum of a \emph{XRISM} pointing in the center of this cluster, based on our model with turbulent and rotating ICM. 

The paper is organized as follows: in Sects.~\ref{sec:data} and \ref{sec:models}, we describe, respectively, the observational data and the models; Sect.~\ref{sec:statistic} details the statistical method used in our analysis; in Sect.~\ref{sec:results}, we present the results of our analysis, we compare them with those of previous works, and we show and discuss the mock spectrum of A2255; 
Sect.~\ref{sec:conclusion} summarizes our main findings.
Throughout this work, we assume a flat $\Lambda$CDM cosmological model\footnote{We assume the flat $\Lambda$CDM model as implemented in the python library Astropy \citep{astropy18}.}, with present-day matter density parameter $\Omega_{\mathrm{m},0}=0.3$ and Hubble constant $H_0=70\,\kms \mathrm{Mpc^{-1}}$.  We indicate with $M_\Delta$ the mass  enclosed within a sphere, centred in the cluster centre, with a radius $r_\Delta$ such that the average density in the sphere is $\Delta$ times the critical density of the Universe $\rhocrit\equiv3H^2(z)/(8 \pi G)$, where $H(z)$ is the Hubble parameter.
In the Bayesian analysis, we adopt the 68\% Highest Posterior Density Interval (HPDI\footnote{The HDPI is estimated using the Arviz python library \citep{arviz19}.}) as the $1\sigma$ credible interval of the marginal posterior \citep[e.g., section~2.3 of][]{Gelman13}.



\section{Published data from X-ray, SZ and WL analyses of X-COP clusters}
\label{sec:data}
In this work, we analyze the clusters belonging to the sample of the XMM-Newton Cluster Outskirts Project (X-COP), a large observational campaign designed to advance our understanding of the virialization region of galaxy clusters \citep{X-COP}.
The X-COP sample consists of 12 nearby ($0.04 < z < 0.09$), massive ($2 \times 10^{14}\,\Msol \lesssim \Mfive \lesssim 3\times 10^{15}\, \Msol$) galaxy clusters. 
The clusters were selected to have the highest signal-to-noise ratio of the SZ effect as resolved in the \textit{Planck} maps \citep{P14}.
The cluster sample was further refined by excluding the objects that appeared significantly disturbed. 
The remaining clusters were followed up with XMM-\textit{Newton} to obtain detailed spectral data out to $\sim \rfive$.
As discussed in \cite{Eckert22}, in clusters with a very peaked density profile near the cluster center, the Point Spread Function (PSF) of XMM-\textit{Newton} contaminates the spectra extracted from a given radial bin with some emission coming from neighboring bins.
Given that in the model-data comparison we neglect this effect, in this work we exclude A2029, which has a very peaked central density profile. Thus our sample of observed clusters consists of the 11 objects listed in Table~\ref{tab.mass}.



\subsection{ICM thermodynamic properties from X-rays}
\label{sec:Xrays}
We present here the properties of the X-COP clusters derived from the X-ray spectral analysis of \citet[][hereafter \citetalias{Ghirardini19}]{Ghirardini19}  and available at the X-COP website\footnote{\url{https://dominiqueeckert.wixsite.com/xcop}}.

For each cluster, $\Nbins$ X-ray spectra were extracted from the relative XMM-{\it Newton} mosaic: the $i$-th spectrum is taken in a circular annulus with inner and outer radii $\Ri$ and $\Rplus$, respectively, assuming as center the peak of the X-ray surface brightness map and out to approximately $\rfive$
\citepalias[see][for details]{Ghirardini19}.  
\citetalias{Ghirardini19} fitted the $i$-th spectrum with the plasma emission code {\tt apec} (Astrophysical Plasma Emission Code\footnote{\url{https://heasarc.gsfc.nasa.gov/xanadu/xspec/manual/XSmodelApec.html}}; \citealt{Smith01}), as implemented in the software for the spectral analysis {\tt XSPEC} \citep{Xspec}.
For our study, the main results of each spectral analysis are the normalization of the thermal emission $Norm$ and the spectroscopic temperature $\Tsp$.
In the model-data comparison, we excluded the bins covering a region partly or totally within 50 kpc from the cluster center, where the presence of a central brightest cluster galaxy (BCG) and the interplay between gas cooling and heating, not accounted for in our models, could be relevant.

We adopt a Cartesian reference system with origin in the cluster center, where $\vec{r} = (x, y, z)$ is the position vector in the three-dimensional (3D) space, such that the \textit{x}-axis is along the line of sight.
In the {\tt apec} emission code, $Norm$ is a function of the 3D density distribution of the plasma. In the $i$-th cylindrical shell, with radii $\Ri$ and $\Rplus$, and having the $x$-axis as symmetry axis, we can write it as
\begin{equation}
    \label{eq.Norm}
    \Normi := C \int_{0}^{2\pi} \int_{\Ri}^{\Rplus} \left(2 \int_{0}^{\infty} \nH \nee \diff x \right) \Rhat \diff \Rhat \diff \phihat,
\end{equation}
where $\phihat$ is the azimuthal angle in the plane of the sky, $\nee(\vec{r})$ and $\nH(\vec{r})$ are the electron number density of the ICM and the equivalent number density of the hydrogen nuclei, respectively, and $C=10^{-14} / \left [ 4\pi \Dang^2 (1+z_0)^2 \right]$, with $\Dang$ and $z_0$ the angular distance and systemic redshift of the cluster, respectively, is the normalization factor adopted in {\tt XSPEC}.
In our analysis we will compare our models with the quantity
\begin{equation}
    \label{eq.NNN}
    \Niii = \frac{\Normi}{\pi \left[ \left(\Rplus / [\arcminute] \right)^2 - \left(\Ri / [\arcminute] \right)^2 \right]},
\end{equation}
whose values are provided by \citetalias{Ghirardini19}.


\subsection{ICM thermodynamic properties from SZ effect}
\label{sec:SZ}
We present here the properties of the X-COP clusters as inferred from the SZ effect signal in the \textit{Planck} maps by \citetalias{Ghirardini19} and reported in the X-COP website.

\citetalias{Ghirardini19} analyzed the Compton $y$-parameter maps, under the assumption that the relation between the "SZ pressure", $\PSZ$, and the $y$-parameter in equation 3 of \citetalias{Ghirardini19} holds.
Radial profiles were extracted in $\NSZ$ circular bins, and then deconvolved by the \textit{Planck} PSF and geometrically deprojected under the assumption of spherical symmetry.
It follows that they recovered the SZ pressure as the average in the $j$-th spherical shell with radii $\radj$ and $\radplus$, $\PSZj$, where $j = \{1, ..., \NSZ\}$ \citepalias[see Sect. 2.5 of][for details]{Ghirardini19}.
In this analysis, a covariant matrix accounts for the errors and the cross-correlations between the values of $\PSZj$ at all $j$. 
Given that the diagonal terms of the covariance matrix are dominant, we neglect the off-diagonal terms and consider only the errors from the diagonal terms in the model-data comparison.
Given that the size of the three central radial bins in $\PSZj$, as widely discussed in \citetalias{Ghirardini19}, are smaller than the Point Spread Function of \textit{Planck}, we excluded them from the model-data comparison.
The used $\PSZ$ data range from $\sim 0.5 \rfive$ out to $\sim 3 \rfive$ in all the clusters.

\subsection{Cluster mass distribution}
\label{sec:mass}
The spherically-averaged mass density profile of the observed clusters and those formed in cosmological simulations are routinely described by the virial mass $\Mtwo$ and by the concentration $c_{200} = \rtwo / r_{-2}$, where $r_{-2}$ is the radius at which the logarithmic slope of the density profile is $-2$.

We take the distribution of $\Mtwo$ measured from WL analysis, when available, as reference estimate of the cluster mass.
Specifically, we take from column 4 of table A2 of \cite{Herbonnet20} the average and the $1\sigma$ scatter of WL mass estimates for the clusters A85, A1785, A2142 and Zw1215.
These values defined our a-priori normal distribution of $\Mtwo$ for each of these clusters.
For the remaining clusters, we assumed the a-priori normal distribution of $\Mtwo$ as follows. 
We take the median of the $\Mtwo$ from table~2 in \cite{Eckert19}, who estimated the $\Mtwo$ distributions of each cluster of the X-COP sample using a correction for the hydrostatic masses based on the assumption of the universal baryon fraction.
We conservatively adopt as standard deviation (normalized to the median $\Mtwo$) $\sigma=\left(\sigma_{\rm ubf}^2+\sigma_{\rm WL}^2\right)^{1/2}$, where  $\sigma_{\rm ubf}$  is (for each cluster) the average between the upper and lower relative uncertainties on $\Mtwo$ (taking the data from table 2 of \citealt{Eckert19}), and $\sigma_{\rm WL}$ is the average relative uncertainty on WL estimates of $\Mtwo$ for the aforementioned five clusters with WL measurements.

We assumed a Gaussian prior distribution of $\ctwo$.
For each cluster the median $\ctwo$ is obtained from the median $\Mtwo$, using the mass-concentration relation in equation 5 of \cite{Ragagnin21} (with coefficients given by their equation 7 evaluated for the cosmological parameters adopted in this work). 
Following \cite{Ragagnin21}, for all clusters we assume 0.38 as standard deviation of the prior distribution of $\ctwo$.


\section{Models for a pressure-supported rotating ICM}
\label{sec:models}

We present here two families of axisymmetric cluster models with rotating ICM that we applied to our sample of clusters: in one family the ICM is assumed to have no turbulence, while in the other the turbulence of the ICM is accounted for.
Models without turbulence, though idealized, are interesting, because they allow us to estimate the maximum room for rotation in the ICM, and could be not too unrealistic, if the negligible turbulence support, measured in the central regions of the Perseus cluster and A2029 \citep{Hitomi16, XRISM_A2029_1}, is confirmed at larger distances from the center and in other clusters. 

Both families belong to the class of models presented by \citetalias{Bartalesi24} \citep[see also][]{Bianconi13,Nipoti15}, in which a pressure-supported and rotating ICM has a composite polytropic distribution \citep[e.g.][]{Cur00} 
 and is in equilibrium in the gravitational potential of a dark-matter halo (the self-gravity of the gas is neglected). While \citetalias{Bartalesi24} considered also oblate and prolate dark halos, here, for simplicity, we limit ourselves to the case of spherical dark halos: in particular we adopt a spherical Navarro-Frenk-White \citep[NFW;][]{NFW96} gravitational potential, which is fully determined by the parameters $\Mtwo$ and $c_{200}$ \citepalias[see e.g.\ section 2.2 of][]{Bartalesi24}.
We generalize the formalism of \citetalias{Bartalesi24} by interpreting the gas pressure $p$ as
\begin{equation}
p=n\kboltz \Tgas+\pturb,    
\end{equation}
where $n$ is the gas number density and $\pturb$ is the turbulent pressure, which we parameterize as $\pturb=\alphaturb p$, with $\alphaturb=\alphaturb(p)$ a dimensionless quantity in the range $0\leq \alphaturb<1$, representing the fraction of turbulent pressure support. The ICM temperature is thus given by 
\begin{equation}
\Tgas=(1-\alphaturb)\frac{p}{n\kboltz}.
\end{equation}

\subsection{Intrinsic properties of the models}
\label{sec:intrinsic}

We describe the intrinsic properties of the models adopting a cylindrical coordinate system $(R,\phi,z)$. Given that the gas distribution is barotropic, the gas rotation velocity $u_\phi$ depends only on $R$: as in 
\citetalias{Bartalesi24},
we assume
\begin{equation}
\label{eq.uphi}
u_{\phi}(R)=4 \upeak \frac{S}{(1+S)^2},
\end{equation}
where $S\equiv R/\Rpeak$, $\Rpeak$ is the radius at which $u_{\phi}$ is maximum and\footnote{In \citetalias{Bartalesi24} the same rotation law is expressed in terms of the quantities $u_0 = 4 \upeak$ and $R_0 = \Rpeak$.} $\upeak \equiv u_\phi(\Rpeak)$.  
This functional form of $u_\phi(R)$ is relatively flexible and is able to reproduce the average rotation velocity profiles of the ICM found in cosmological simulations for clusters with masses comparable to those of X-COP clusters \citepalias[see][]{Bartalesi24}.

In the composite polytropic models here adopted, the gas number density 
distribution is 
\begin{equation}
    \label{eq.density}
    n(R,z)=\nstar \left[1- \frac{\gamma'-1}{\gamma'} \frac{\nstar \mu \proton}{\ptotstar} \Delta \Phieff(R,z)\right]^{\frac{1}{\gamma'-1}}
\end{equation}
and the pressure distribution is
\begin{equation}
    \label{eq.pressure}
    \ptot(R,z)=\ptotstar \left(\frac{n(R,z)}{\nstar} \right)^{\gamma'},
\end{equation}
with $\gamma'=\gamIN$ and $\gamma'=\gamOUT$ in the inner and outer regions of the cluster, respectively. The effective potential $\Phieff$, depending on $\Mtwo$, $c_{200}$, $\upeak$ and $\Rpeak$, is defined in equations 24 - 25 of \citet{Bartalesi24}. In the above equations, $\nstar = n(\Rbreak, 0)$ and $\ptotstar = \ptot(\Rbreak, 0)$, with $\Rbreak$ the radius in the equatorial plane where the value of the polytropic index changes; $\mu$ is the mean molecular weight and $\proton$ is the proton mass.
We assume position-independent metallicity of 0.3 times the solar value, for which $\mu = 0.6$ and $n / \nee = 1.94$, or $\nee = 1.17 \nH$ in a fully ionized plasma.
For models both with and without turbulence, we found convenient defining the parameter $T_\star\equiv\ptotstar/(\nstar\kboltz)$, which is the gas temperature at $(\Rbreak,0)$ when $\alphaturb=0$ and can be interpreted as an equivalent gas temperature  at $(\Rbreak,0)$ when $\alphaturb\neq 0$.

In models with turbulent ICM  ($0\leq \alphaturb < 1$) we adopt  the following law to describe the dependence of $\alphaturb$ on $\ptot$:
\begin{equation}
\label{eq.alpha_turb}
\alphaturb(\ptot) = (\alphainf - \alphazero) \frac{\ln \left[1 + \ptot / (\xi p_\star) \right]}{\ptot / (\xi p_\star)} + \alphazero,
\end{equation}
where $\xi$ is a dimensionless parameter.  
In Eq.~(\ref{eq.alpha_turb}), $\alphazero$ and $\alphainf$ are the asymptotic values of $\alphaturb$ for $\ptot \gg \xi p_\star$ and $\ptot \ll \xi p_\star$, respectively (the transition between the two regimes occurs where the pressure has values around  $\xi p_\star$). 
Given that, for a pressure-supported cluster, the pressure decreases outward, $\alphaturb$ increases outward if $\alphainf > \alphazero$.
The contribution to the ICM support against cluster gravity from turbulent motions is routinely measured as a spherically averaged $\alphaturb$ profile in clusters formed in cosmological simulations. 
Using a multiscale filtering technique \cite[see][]{Vazza12}, which is believed to isolate the uncorrelated component of the velocity field of the ICM, \cite{angelinelli20} \citepalias[hereafter][]{angelinelli20} accurately estimated the profiles of $\alphaturb$ for a sample of 68 clusters formed in a cosmological, non-radiative simulation, without AGN and stellar feedback.
They, then, presented a functional form in their equation 11 that successfully reproduces the outward increasing radial dependence of the median $\alphaturb$ profile of these clusters.
The functional form of $\alphaturb$ presented in eq. (\ref{eq.alpha_turb}) is thought to generalize that of \citetalias{angelinelli20} to non-spherical analyses, in particular to axisymmetric systems for the scope of this work. 


In summary, the family of models without turbulence ($\alphaturb=0$) has 9 free parameters:
 $\gamIN$, $\gamOUT$, $\nstar$, $T_\star$, $\Rbreak$, $\Mtwo$, $c_{200}$, $\Rpeak$ and $\upeak$.
The family of models with turbulence has 12 free parameters: $\alphainf$, $\alphazero$ and $\xi$, in addition to the nine parameters of the other family.

\subsection{From models to observational data}
\label{sec:mod2data}

We define here the quantities, obtained from the intrinsic properties of our models (see Sect.~\ref{sec:intrinsic}), that can be compared with the observational data detailed in Sect.~\ref{sec:data}.
We use the Cartesian reference system and the radial grids $\Ri$ and $\radj$ introduced in Sect.~\ref{sec:data}.

In the {\tt apec} emission code, $\Tsp$ is formally the temperature of an isothermal emitting plasma (see Sect.~\ref{sec:Xrays}). 
Given that the ICM in our model is multi-temperature, we adopted, as an approximation of $\Tsp$, the spectroscopic-like temperature $\Tsl$ \citep[see][for details]{M04}, such that we will compare $\Tsl$ derived from our models to $\Tsp$. 
$\Tsl$ in the $i$-th cylindrical shell, with radii $\Ri$ and $\Rplus$, is
\begin{equation}
    \label{eq.Tsl}
    \Tsli=\frac{\int_{0}^{2\pi}
    \int_{\Ri}^{\Rplus} \int_{0}^{\infty} \nH \nee \Tgas^{1/4}  \Rhat \diff x \diff \Rhat \diff \phihat} {\int_{0}^{2\pi}
    \int_{\Ri}^{\Rplus} \int_{0}^{\infty} \nH \nee \Tgas^{-3/4}  \Rhat \diff x \diff \Rhat \diff \phihat},
\end{equation}
where $\Tgas(\vec{r})$ is the intrinsic temperature of the ICM.

The SZ pressure, obtained from the intrinsic properties of the ICM in our models, is assumed to be the average of the thermal pressure distribution of the ICM in the $j$-th spherical shell, with radii $\radj$ and $\radplus$,
\begin{equation}
    \label{eq.p_e}
    \PSZj = \frac{3 \int_{\radj}^{\radplus} \nee \kboltz \Tgas r^2 \diff r}{\radplus^3 - \radj^3},
\end{equation}
where 
$r$ is the spherical radius and $\kboltz$ the Boltzmann constant.

In the comparison between the models and the observed dataset, we assumed that our axisymmetric models are observed edge-on, that is with line of sight $x$ orthogonal to the symmetry and rotation axis $z$. 
As we have seen in Sect.~\ref{sec:data}, the observational data we want to compare with are measurements of quantities in either circular annuli or in spherical shells centered in the cluster center. 
Though we could perform analogous measurements for our models starting from Eqs.~(\ref{eq.Norm}), (\ref{eq.Tsl}) and (\ref{eq.p_e}), in order to save computational time in the model-data comparison, we approximated such measurements with estimates obtained by considering only properties of the models in the equatorial plane ($z=0$).
The details of this approximation, which turns out to be sufficiently good for our purposes, are given in Appendix \ref{sec:approx}. 
In particular, imposing that the gas density is zero when $|x|>\rout$ or $|y|>\rout$, where $\rout$ is the outer radius of the outhermost spherical shell, we applied the approximations (\ref{eq.appXraynew}) and (\ref{eq.appSZ}) to our models.
Thus, Eqs.\ (\ref{eq.Norm}) and (\ref{eq.Tsl}) become, respectively, 
\begin{equation}
    \label{eq.model_Norm}
   \Normi = 4\pi C \int_{\Ri}^{\Rplus} y \diff y \left( \int_{0}^{\rout} \diff x\, \nH(R, 0) \nee(R, 0)  \right)
\end{equation}
and
\begin{equation}
    \label{eq.model_Tsl}
    \Tsli =\frac{
    \int_{\Ri}^{\Rplus} y \diff y \left( \int_{0}^{\rout} \diff x\,\nH(R, 0) \nee(R, 0) \Tgas^{1/4}(R, 0) \right)   } {
    \int_{\Ri}^{\Rplus} y \diff y \left( \int_{0}^{\rout} \diff x\,\nH(R, 0) \nee(R, 0) \Tgas^{-3/4}(R, 0) \right) },
\end{equation}
with $R=\sqrt{x^2+y^2}$; Eq.\ (\ref{eq.p_e}) becomes
\begin{equation}
    \label{eq.model_press}
    \PSZj = \eta \frac{3 \int_{\radj}^{\radplus} y^2 \diff y\, \nee(|y|, 0) \kboltz \Tgas(|y|, 0) }{\radplus^3 - \radj^3},
\end{equation}
where $\eta$ is a parameter accounting for the possible systematic offset between the electron pressure of the ICM and that measured from the Compton $y$-parameter maps (see Appendix \ref{sec:systematics}).

\begin{table}
      \caption[]{Priors on $\Mtwo$ and $\ctwo$ of our models.}
         $$ 
         \begin{array}{p{0.2\linewidth}llll}
            \hline
            \noalign{\smallskip}
            Cluster & z & \Mtwo^{\mathrm{median}} & \sigma_{\mathrm{M}} & \ln c_{200}^{\mathrm{median}} \\
            \noalign{\smallskip}
            \hline
            \noalign{\smallskip}
            A85 & 0.0555 & 0.84 & 0.27 & 1.22 \\
            A644 & 0.0704 & 0.83 & 0.25 & 1.24 \\
            A1644 & 0.0473 & 0.66 & 0.20 & 1.24 \\
            A1795 & 0.0622 & 1.39 & 0.28 & 1.19 \\
            A2142 & 0.090 & 1.45 & 0.30 & 1.19 \\
            A2255 & 0.0809 & 1.07 & 0.32 & 1.21 \\
            A2319 & 0.0557 & 2.01 & 0.59 & 1.17 \\
            A3158 & 0.0597 & 0.73 & 0.22 & 1.23 \\
            A3266 & 0.0589 & 1.45 & 0.49 & 1.19 \\
            RXC1825 & 0.0650 & 0.69 & 0.20 & 1.24 \\
            Zw1215 & 0.0767 & 0.51 & 0.27 & 1.26 \\
            \noalign{\smallskip}
            \hline
         \end{array}
         $$
     \tablefoot{Columns: cluster name, redshift, median and standard deviation of the normal prior on $\Mtwo$, median of the prior on $\ctwo$. The standard deviation of $\ctwo$ is 0.38 for all the clusters. $\Mtwo^{\mathrm{median}}$ and $\sigma_{\mathrm{M}}$ are in units of $10^{15}\, \Msol$.
        }
     \label{tab.mass}
\end{table}

\begin{table}
      \caption[]{Lower and upper bounds of the uniform prior distributions of the MCMC parameters $\gamIN$,
            $\gamOUT$, 
            $\nu_\star$, 
            $T_\star$,
            $\Rbreak$, 
            $\log \fff$, 
            $\log \gggg $,
            $\upeak$, 
            $\Rpeak$, $\eta$ and $\log \sss$  (in common to models both with and without turbulence), and  of the parameters
            $\log \xi$, 
            $\alphainf$, 
            $\alphazero$ (used only in the model with turbulence).
      }
     $$ 
         \begin{array}{ll}
            \hline
            \noalign{\smallskip}
            \mathrm{Parameter} & \mathrm{Range}\\
            \noalign{\smallskip}
            \hline
            \noalign{\smallskip}
            \gamIN & [0.6, 1.6]\\
            \gamOUT & [0.9, 1.6]\\
            \nu_\star & [-1.0, 2.0]\\
            T_\star /\mathrm{keV} & [3.0, 18.0]\\
            \Rbreak /\mathrm{kpc} & [50, 0.35 (\rtwobar/\mathrm{kpc})]\\
            \log \fff & [-2.0, 1.7]\\
            \log \gggg & [-2.5, 1.3]\\
            \upeak / (\kms) & [50, 2000]\\
            \Rpeak / \mathrm{kpc} & [50, (\rout/\mathrm{kpc})]\\
            \eta & [0.3, 1.8]\\
            \log \sss & [-2.5, 1.0]\\
            \log \xi & [-1.5, 0]\\
            \alphainf & [0, 0.95]\\
            \alphazero & [0, \alphargs]\\
            \noalign{\smallskip}
            \hline
         \end{array}
     $$ 
     \tablefoot{First column: parameter. Second column: lower and upper bounds of the uniform prior distribution.
The upper bound on $\Rbreak$ is motivated by the fact that no cluster is known to have cooling radius larger than $0.35 \rtwo$ \citep[see, e.g.,][]{Cavagnolo09}. Here $\rtwobar = 3 \Mtwobar / \left[800 \pi \rhocrit \right]^{1/3}$, where $\Mtwobar$ is the median of the prior on $\Mtwo$. 
The prior distributions of the parameters $\Mtwo$ and $c_{200}$, not listed here, are defined in Sect.~\ref{sec:prior} and reported in Table \ref{tab.mass}. 
     }
     \label{tab.prior}
\end{table}

\section{The statistical method}
\label{sec:statistic}
For each cluster of our sample we computed both $\alphaturb = 0$ and $\alphaturb \ge 0$ models. 
Taking mass distributions based on available estimates (see Sect.~\ref{sec:mass}), these models, presented in Sect.~\ref{sec:models}, are tuned to reproduce the radial profiles of $\NNN$, $\Tsp$ and $\PSZ$ as resolved in the XMM-\textit{Newton} and \textit{Planck} maps (see Sects.~\ref{sec:Xrays} and \ref{sec:SZ}) via a Markov Chain Monte Carlo (MCMC) method.
This section details the statistical method: we present the form of the likelihood and the prior distributions in Sects.~\ref{sec:likelihood} and \ref{sec:prior}, respectively, and the method used to estimate the posterior distribution of the model in Sect.~\ref{sec:posterior}.
The details of the MCMC algorithm are described in Appendix~\ref{sec:MCMC}.

\subsection{Likelihood}
\label{sec:likelihood}
Using the models described in Sect.~\ref{sec:models}, we reconstruct the intrinsic properties of the ICM (density, turbulent and thermal pressures, and rotation speed) and the gravitational potential in which it is in equilibrium.
From the model of the intrinsic quantities of the ICM,  we evaluated the radial profile of $\NNN$, $\Tsl$ and $\PSZ$ as described in Sect.~\ref{sec:mod2data} (we recall that the normalization of the $\PSZ$ profile is regulated by $\eta$, which is one of the free parameters of the MCMC).
The statistical errors, $\sigmastat$, on the observed $\NNN$, $\Tsp$ and $\PSZ$ are obtained as the average between the upper and lower $1 \sigma$ errors reported in the data analyses.
We consider three parameters, $\fff$, $\gggg$ and $\sss$, to account for a systematic contribution, $\epsilon \sigmastat$ with $\epsilon = \{ \fff, \gggg, \sss \}$, to the uncertainty in the values of $\NNN$, $\Tsp$ and $\PSZ$, respectively.  
$\fff$, $\gggg$ and $\sss$ are assumed to have the same value in all the radial bins.
Specifically, we assumed each datum of $\NNN$, $\Tsp$ and $\PSZ$ to be randomly generated from a normal distribution with a median equal to the corresponding $\NNN$, $\Tsl$ and $\PSZ$ in the model and a standard deviation equal to $\sqrt{1 + \epsilon^2}  \sigmastat$.
It follows that the generic likelihood of each datum $Q_{m,i_m}$, $\mathcal L (Q_{m,i_{m}})$, where $Q_{m}$ with $m = \{1, 2, 3\}$ indicates $\NNN$, $\Tsp$ or $\PSZ$, respectively, and $i_{m}$ the radial bin in which $Q_{m}$ was extracted, is a normal distribution with the same median and standard deviation as described above.
The radial bins used in the fitting are defined in Sects.~\ref{sec:Xrays} and \ref{sec:SZ}.

\subsection{Prior distributions}
\label{sec:prior}

Let us consider the marginal prior of the generic parameter $\theta_{h}$, $\mathcal \pi(\theta_{h})$, where $h = \{1, ..., H\}$ with $H$ the number of the parameters in the fitting.
We found convenient using as MCMC parameter $\nu_\star \equiv \log [\nstar / (\mathrm{10^{-3} cm^{-3}})]$ instead of $\nstar$.
In addition to the parameters of the models (see Sect. \ref{sec:intrinsic}), for the joint analysis of the X-ray and SZ data we introduced the four parameters $\eta$, $\fff$, $\gggg$ and $\sss$ (see Sect. \ref{sec:likelihood}), so  for the $\alphaturb = 0$ models we have  $H = 13$ parameters ($\gamIN$, $\gamOUT$, 
            $\nu_\star$, 
            $T_\star$,
            $\Rbreak$, 
            $\log \fff$, 
            $\log \gggg $,
            $\upeak$, 
            $\Rpeak$, $\eta$, $\log \sss$, $\Mtwo$ and $\ctwo$) while for the $\alphaturb \ge 0$ models we have $H = 16$ parameters 
            ($\log\xi$, 
            $\alphainf$ and  
            $\alphazero$, in addition to those of the $\alphaturb=0$ models).
            
            We assume $\pi(\Mtwo)$ and $\pi(\ctwo)$ to be Gaussian, with the medians and standard deviations reported for each cluster in Table~\ref{tab.mass} (see also Sect. \ref{sec:mass}).
For all the other parameters we assume a uniform prior distribution, with upper and lower bounds reported in Table~\ref{tab.prior}. 
The choice of the upper and lower bounds of the parameters regulating the turbulent support ($\log\xi$, $\alphazero$ and $\alphainf$) requires a brief comment. 
The parameter $\log\xi$ sets the distance from the center where there is a transition between an inner region ($p\gg \xi p_\star$) in which $\alphaturb\to\alphazero$ and an outer region  ($p\ll \xi p_\star$) in which $\alphaturb\to\alphainf$. 
We adopted a prior on $\log\xi$ uniform in the range [-1.5, 0]. 
This is motivated by the expectation (based on the properties of the model SRM of \citetalias{Bartalesi24}) that values of $\log\xi$ in the range [-1.5, 0] map the radial range (0.15 - 1)$\rtwo $, which excludes to have the transition within the cluster core ($\log \xi = 0$ sets the transition in the $\alphaturb$ profile at $R \approx \Rbreak$ in the equatorial plane).
 For our models applied to the X-COP clusters, a turbulent velocity dispersion higher than $\sigmargs = 500 \, \kms$ at a distance smaller than $\approx$ 150 kpc would violate the 90\% upper limit on the broadening of the X-ray emitting lines from \textit{RGS} data.
 We thus set the upper bound of $\pi(\alphazero)$ to $\alphargs = \sigmargs^2/\left[\sigmargs^2+\kboltz T_0 / (\mu \proton)\right]$, where $T_0$ is the best-fit observed spectroscopic temperature in the innermost annulus of each cluster \citepalias[taken from][]{Ghirardini19}.
$\alphainf$ accounts for the turbulent support at large radii, which is observationally unconstrained. We thus assume 0.95 as upper bound of $\pi(\alphainf)$, which corresponds to dominant turbulent pressure in the outskirts (see Sect.~\ref{sec:intrinsic}).
We set the lower bounds of $\pi(\alphazero)$ and $\pi(\alphainf)$ to 0, to conservatively include the possibility of negligible turbulence.



\begin{figure*}
   \centering
   \includegraphics[width=0.49\textwidth]{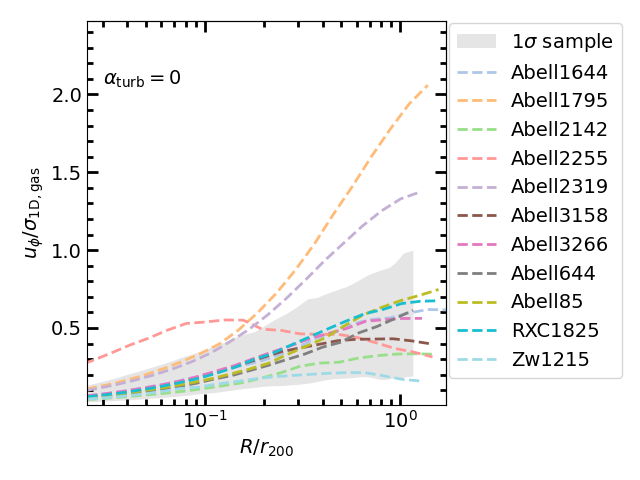}
\includegraphics[width=0.49\textwidth]{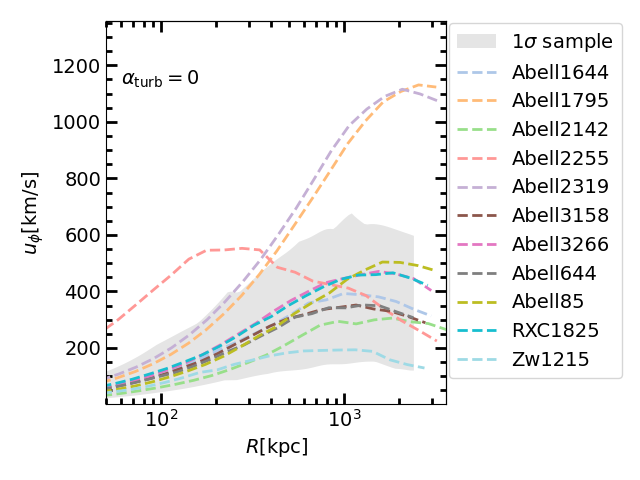}
    \caption{
    Equatorial plane velocity profiles of the ICM in the $\alphaturb=0$ models of the clusters of our sample. Left panel: ratio between gas rotation velocity and velocity dispersion (measured in the equatorial plane) as a function of  the cylindrical radius normalized to the virial radius $\rtwo = 3 \Mtwo / \left[800 \pi \rhocrit \right]^{1/3}$ computed using the median value of $\Mtwo$ for each cluster.  
    Right panel:  gas rotation velocity as a function of cylindrical radius in physical units. 
    In each panel the dashed lines are the median profiles of the individual clusters (as indicated in the legend), whereas the gray band represents the interval between the 16th and 84th percentiles of the distribution of the profiles of the entire sample of clusters. 
    }
    \label{fig.sample_rot}
\end{figure*}

\begin{figure*}
   \centering
   \includegraphics[width=0.49\textwidth]{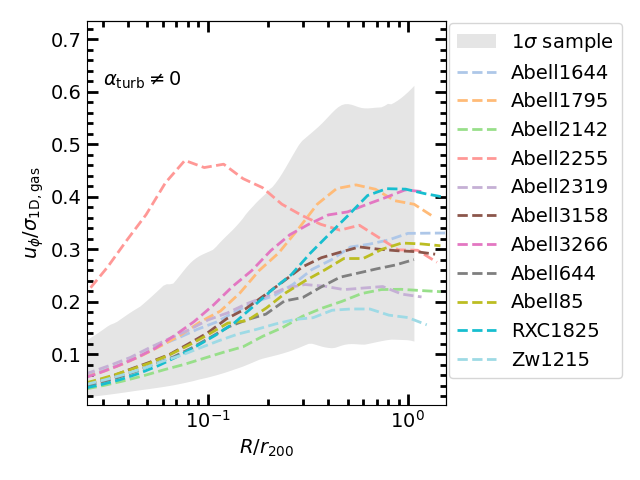}
   \includegraphics[width=0.49\textwidth]{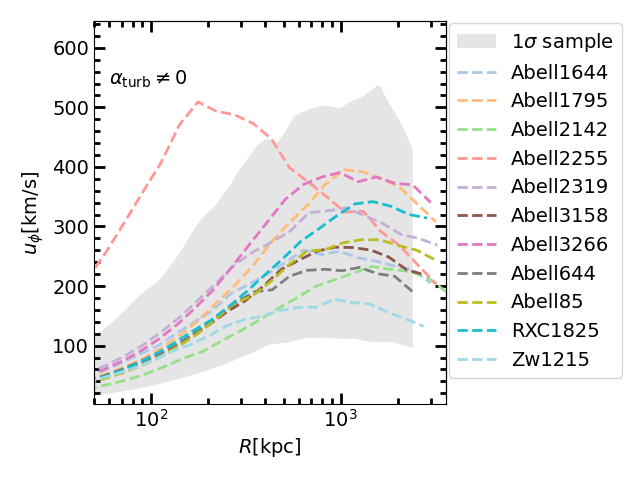}  
 \includegraphics[width=0.49\textwidth]{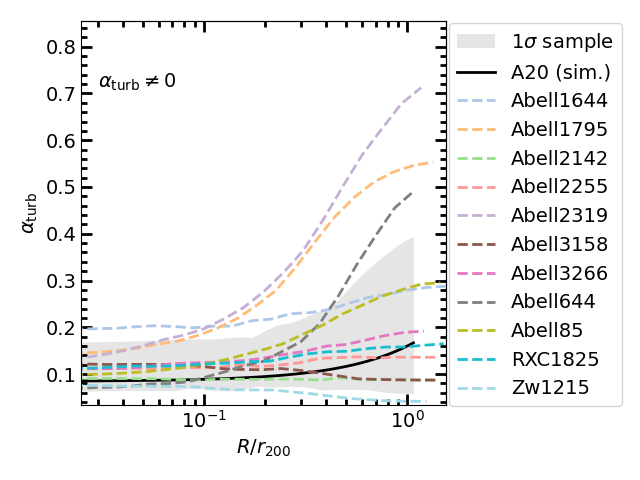}
   \includegraphics[width=0.49\textwidth]{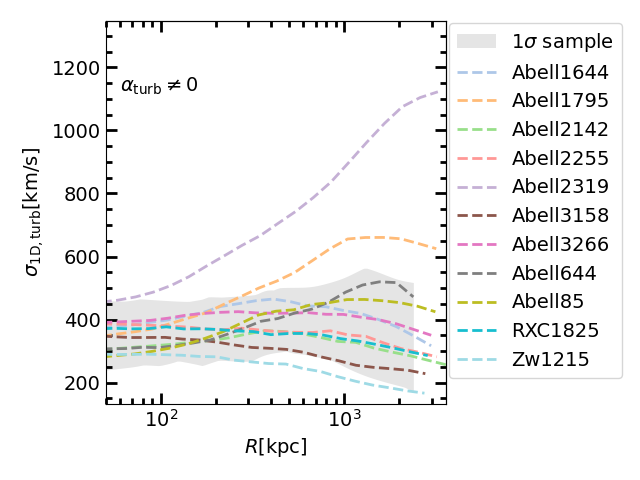}
    \caption{
    Equatorial plane rotation and turbulence profiles of the ICM in the $\alphaturb\neq 0$ models of the clusters of our sample. 
    Upper panels.
    Same as Fig.~\ref{fig.sample_rot}, but for models with $\alphaturb \ge 0$. 
    Lower left panel. Median turbulent-to-total pressure ratio $\alphaturb$ as a function of cylindrical radius normalized to the virial radius $\rtwo$ (see caption of Fig.~\ref{fig.sample_rot}) for the same models as in the upper panels. Lower right panel. 1D turbulent velocity dispersion as a function of the cylindrical radius in physical units for the same models as in the upper panels. 
    In each panel the color coding and the meaning of the gray band are the same as in Fig.\ \ref{fig.sample_rot}.
    In the lower right panel, we overplot the $\alphaturb$ profile measured in simulations by \citetalias{angelinelli20} (black solid line; the profile was taken from their equation~11, where the best-fit value of $a_0$, $a_1$ and $a_2$ came from the first row of their Table~1, and assuming $r_\mathrm{200,m} / \rtwo \approx 1.67$).
    }
    \label{fig.sample_turb}
\end{figure*}

\subsection{Posterior distributions}
\label{sec:posterior}
According to Bayes' theorem, the posterior distribution of the parameter vector $\vec \theta = (\theta_1, ..., \theta_H)$, $\Prob(\vec \theta)$, is obtained from $\ln \Prob = \sum_{h = 1}^{H} \mathcal \ln \left[\pi(\theta_{h})\right] + \sum_{m = 1}^{3} \sum_{i_{m}} \ln \left[\mathcal L (Q_{i_m})\right]$, where $\pi(\theta_{h})$ and  $\mathcal L(Q_{m})$ are described in Sects. \ref{sec:prior} and \ref{sec:likelihood}, respectively. 
The MCMC algorithm individually maximizes $\ln \Prob$ of each cluster.
When it converges to a stationary posterior, we obtain the discrete posterior consisting of a sample of the parameter vector (see Appendix ~\ref{sec:MCMC}, for details). 
The relative frequency of the values of the parameters in this sample can be visualized in corner plots, which we present for the parameters $\Mtwo, \ctwo, \upeak, \Rpeak$ and $\alphazero$ (if $\alphaturb \neq 0$) for all clusters in Appendix~\ref{sec:profiles}.

We chose to characterize the marginal posterior distributions using the 68\% HPDI as a credible interval (see Sect.~\ref{sec:intro}), which is very similar to the interval between the 16th and 84th percentiles for symmetric distributions, but tends to be more meaningful than the 16th and 84th percentiles for highly skewed distributions. For the sake of conciseness, in the following we refer to the lower and upper bounds of the 68\% HPDI credible interval on the marginal posterior distribution on a parameter, simply as upper and lower limit on that parameter.


For each parameter vector belonging to our sampling of the posterior distribution (see above), we computed a few relevant quantities for each cluster, needed for the interpretation of the results presented in Sect. \ref{sec:results}.
In particular, we derived the profiles of $\NNN$, $\Tsl$ and $\PSZ$ as described in Sect. \ref{sec:mod2data}, and the rotation curve using Eq.~(\ref{eq.uphi}).
The 1D velocity dispersion of the gas particles at a given point in the equatorial plane is given by $\sigmagas(R, 0) = \sqrt{p(R, 0) / \left[\mu \proton n(R, 0)\right]}$. 
As a measure of the dynamical importance of rotation, we computed, as a function of radius $R$, the ratio between the gas rotation velocity $u_\phi(R)$ and the gas velocity dispersion $\sigmagas(R,0)$ in the equatorial plane, $u_\phi/\sigmagas$. 
Given that we are interested in the effect of the turbulence when fitting the thermodynamic data with a rotating ICM model, in the $\alphaturb \ge 0$ models,  we also derived the $\alphaturb(p)$ profile through Eq.~(\ref{eq.alpha_turb}), once we have obtained the pressure profile (see Sect.~\ref{sec:intrinsic}). 
The 1D turbulent velocity dispersion in the equatorial plane is then given by $\sigmaturb(R, 0) = \sqrt{\alphaturb(p)} \sigmagas(R, 0)$, where $p=p(R, 0)$.

In Appendix~\ref{sec:profiles}, we show for each cluster the profiles of $\NNN$, $\Tsl$, and $\PSZ$ for the $\alphaturb=0$ model. The $\NNN$, $\Tsl$ and $\PSZ$ profiles of the $\alphaturb \neq 0$ models, not shown, are virtually indistinguishable from the corresponding profiles of the $\alphaturb = 0$ models. 
In the same Appendix, we also show for each cluster the $u_\phi$ profile of the $\alphaturb=0$ model, and the $u_\phi/\sigmagas$ and $\alphaturb$ profiles of the $\alphaturb\geq 0$ model.

\section{Results}
\label{sec:results}

We present here the results obtained by applying our models, presented in Sect.~\ref{sec:models}, to the observed thermodynamic profiles of the X-COP objects through the statistical method described in Sect.~\ref{sec:statistic}, focusing in particular on the radial distribution of the ICM rotation and turbulence.

\subsection{Models of rotating ICM without turbulence}
\label{sec:rotation}


Our models are able to reproduce well the observed properties of the ICM (Sect.~\ref{sec:data}) of all the clusters of our sample from a radius comparable to the BCG size out to $\sim 3 \rfive$, the outermost radial bin in the {\it Planck} $\PSZ$ profile (see Appendix~\ref{sec:profiles}). 
The residuals  between the median of the model and the data are, as a rule, not larger than 6\% in the radial profiles of $\NNN$, $\Tsp$ and $\PSZ$ (the only exceptions are A644 and A2319, for which the residuals are between 6\% and 10\%).
The median value of $\fff$ inferred in the sample is 4.5 (the median $\fff$ in each cluster ranges from 1, in Zw1215, to 10, in A85). 
The values of $\gggg$ and $\sss$ are lower than 1 in all the clusters of the sample, apart from $\gggg$ in A644, A1644 and A2319, whose medians range from 2 to 4, respectively.
The median errors inferred as $\sqrt{1 + \epsilon^2} \sigmastat$ are plotted in the thermodynamic profiles in Appendix \ref{sec:profiles}.

For all the clusters, the posterior distribution $\mathcal P$ of $\upeak$ is unimodal, with a credible interval that encloses the mode (see the corner plots in Appendix \ref{sec:profiles}). 
In most cases, the range between the lower and upper limits of $\upeak$ is wide, implying that the uncertainty on the inferred values of $\upeak$ is large.
The posterior lower limit of $\upeak$ is very close to the lower bound of $\pi(\upeak)$, except for three clusters (A1795, A2255 and A2319).

Considering altogether our sample of objects, we characterize the distribution of the profiles of any quantity, say $q(R)$, by defining 1$\sigma$ lower and upper limits of as follows. 
At any radius $R$ in the equatorial plane, we define the $1\sigma$ lower limit as the median of the 16th percentiles of the distribution of $q$ evaluated at $R$ in each cluster model. The 1$\sigma$ upper limit is defined in the same way, but by taking the 84\% percentiles.  

Figure~\ref{fig.sample_rot} summarizes the rotation properties of our models of the clusters of our sample.
In the left panel, we show the median $u_\phi/\sigmagas$ profiles of the clusters (see Sect.~\ref{sec:posterior}).
A common trend emerges: this population of clusters exhibits a slightly outward-increasing median profile of $u_\phi/\sigmagas$. 
However, the lower limit of the population (lower bound of the gray band in the plot) has everywhere $u_\phi / \sigmagas < 0.2$, corresponding to $\rho u_\phi^2 / p < 0.04$, which indicates a negligible rotation support.
In the right panel, where we show the median profiles of the rotation velocity in the cluster models, the lower limit of the population  (lower bound of the gray band in the plot) corresponds to velocity everywhere below 100 $\kms$.
The population upper limit of the $u_\phi / \sigmagas$ profile (upper bound of the gray band in the left panel) significantly increases outward up to nearly 0.9 (corresponding to a rotation velocity of $\approx$ 600 $\kms$; see right panel).

Taking the difference between the population upper and lower limits as a measure of the 1$\sigma$ scatter, we find that the scatter in $u_\phi / \sigmagas$ and in $u_\phi$ is up to 0.7 and 500 $\kms$, respectively, implying that the determination of the properties of the population in our models is very uncertain.
Three clusters have median $u_\phi / \sigmagas$ and $u_\phi$ profiles significantly higher than the 1$\sigma$ upper limit of the population: A1795 and A2319 have high $u_\phi / \sigmagas$ and $u_\phi$ in the outskirts (with $u_\phi$ higher than 1000 $\kms$), while A2255 at intermediate radii (with $u_\phi \approx 500$ $\kms$).

We conclude that, if turbulence in the ICM is negligible throughout the cluster (which is an hypothesis consistent with the currently available measurements in the central regions of the Perseus cluster and A2029; \citealt{Hitomi16, XRISM_A2029_1}), the current data of the X-COP clusters leave room for an important rotation in the ICM (especially in the three clusters A1795, A2319 and A2255), even if in most cases they are also consistent within the 68\% credible interval with negligible rotation of the ICM.

In the next section, we consider the case in which the contribution of ICM turbulence is accounted for in the models.

\subsection{Models of rotating ICM including turbulence}
\label{sec:turbulence}

When applying the $\alphaturb \ge 0$ model to the clusters of our sample, we find that the posterior distributions of the parameters $\gamIN$, $\gamOUT$, $\nu_\star$, $\Rbreak$, $\log \fff$, $\log \gggg $, $\eta$, $\log \sss$, $\Mtwo$ and $\ctwo$ are very similar to the case of the $\alphaturb=0$ model.
We find instead significant differences in  the posterior distributions of the parameters $T_\star$, $\upeak$ and $\Rpeak$.
This confirms the expectation that the X-ray and SZ data alone are insufficient to distinguish between the two forms of non-thermal support considered in our study.
However, the functional forms adopted for the radial profiles of rotation and turbulence play a pivotal role in separating the rotational contribution to non-thermal support from that from turbulence.
In all the clusters, the upper limit of $\upeak$ for $\alphaturb \ge 0$ tends to be lower than for $\alphaturb = 0$, which indicates that rotation is less important when the ICM is turbulent. This is expected, because turbulence contributes to the pressure support against gravity, thus reducing the room for rotational support.

Figure~\ref{fig.sample_turb} summarizes the properties of the $\alphaturb \ge 0$ models for the clusters of our sample.
As shown in the upper-left panel, when $\alphaturb \neq 0$ the population upper limit on $u_\phi / \sigmagas$ (upper bound of the gray band in the plot) is $\approx 0.6$ (corresponding to a rotation speed of $\approx 500 \,\kms$; see upper-right panel), smaller than $\approx 0.9$ found when $\alphaturb=0$, but still indicating room for significant rotation support, even in the presence of turbulence. 
Conversely, as shown in the lower panels of fig. \ref{fig.sample_turb}, the $1\sigma$ scatter of the turbulent-to-total-pressure ratio ($\alphaturb$, in its left panel) and 1D turbulence velocity dispersion ($\sigmaturb$, in its right panel) in the population ranges from $\approx$ 0.05 and 200 $\kms$ up to 0.4 and 500 $\kms$, characterized by a very significant rise of population upper limit of $\alphaturb$ in the outskirts.

It is particularly interesting to compare $\alphaturb = 0$ and $\alphaturb \ge 0$ models of the clusters whose $\alphaturb = 0$ model shows significant rotation, namely A1795, A2319 and A2255.

The models of A1795 and A2319 have similar properties. The $\alphaturb = 0$ models of both objects have posterior lower limit of $\upeak$ around 850 $\kms$, much higher than the lower bound of the prior of $\upeak$ (see the corner plot in figs. \ref{fig.A1795} and \ref{fig.A2319}, respectively). 
This corresponds to a median $u_\phi / \sigmagas$ exceeding 1 in the outskirts and thus significantly deviating from the average properties of the population (see the right panel of Fig.~\ref{fig.sample_rot}). 
Instead, when $\alphaturb \ge 0$, the models of A1795 and A2319 have posterior lower limit of $\upeak$ coincident with the lower bound of $\pi(\upeak)$ (50 $\kms$).
This translates into a median $u_\phi / \sigmagas$ profile, with a value below 0.4 in the outskirts, which everywhere falls within the 1$\sigma$ interval of the population (see the upper-left panel of Fig.~\ref{fig.sample_turb}).
The lower-left panel of Fig.~\ref{fig.sample_turb} shows the median turbulent-to-total pressure ratio profile $\alphaturb(R)$ of our clusters. 
A1795 and A2319 have $\alphaturb$ significantly increasing outward, with $\alphaturb>0.5$ in the outskirts, standing out from the 1$\sigma$ band of the considered cluster population. 
We can interpret this as a significant shift of the non-thermal support from rotation to turbulence in the outer regions.

In the $\alphaturb \ge 0$ model, A2255 exhibits the highest median $u_\phi / \sigmagas$ profile in a region close to the center, making it the most promising candidate for detecting rotation in the ICM with XRISM/\textit{Resolve}.
The credible interval of $\mathcal P(\upeak)$ in the $\alphaturb \ge 0$ model consists of two separate intervals corresponding to relatively low and high values of $\upeak$, centered at 100 $\kms$ and 600 $\kms$, respectively. 
$\mathcal P(\upeak)$ is relatively flat, from the lower bound of the low-$\upeak$ credible interval (that is the lower bound of the prior) to the upper bound of the high-$\upeak$ credible interval, which is approximately equal to the 1$\sigma$ upper limit of $\upeak$ in the $\alphaturb = 0$ model.
Given the relatively high rotation velocity of the ICM in the inner or intermediate regions, this cluster warrants further investigation through a dedicated XRISM pointing in its center.
In Sect. \ref{sec:A2255}, we propose an observational strategy with XRISM/\textit{Resolve} to validate the rotation peak of 500 $\kms$ at a distance of 200 kpc from the center.


In Appendix \ref{sec:flat}, we compare the flattening of the X-ray surface brightness distribution measured in our models with the measurements of the X-COP clusters.
We found that the $\alphaturb = 0$ models (being more flattened) tend to perform better than the $\alphaturb \ge 0$ models in reproducing the observed distribution of the axial ratios estimated from the X-ray maps. 
This might indicate that the ICM turbulence support is not as high as our models would have inferred.
However, it must be stressed that in our models, where the gravitational potential is spherically symmetric, only the rotation determines the flattening of the X-ray surface brightness distribution. 
The flattening of the ICM can also be influenced by the shape of the dark-matter halo, which can deviate significantly from spherical symmetry  \citep[e.g.][]{Allgood06}. This is widely discussed in \citetalias{Bartalesi24}, where models of the rotating ICM in equilibrium in both prolate and oblate halos were studied. 


\subsection{Comparison with the published studies on the ICM motions}
\label{sec:comparison}

To our knowledge, the present work is the first attempt to reproduce the temperature, density and pressure profiles of the observed clusters with models with rotating and turbulent ICM, consistent with the available mass estimates.
Here we compare our results with published works presenting constraints on the contribution of non-thermal pressure support from observed datasets, where the non-thermal component is totally ascribed to turbulence, and on the turbulent support in clusters formed from cosmological hydrodynamic simulations.

\citet[][hereafter \citetalias{Ettori22}]{Ettori22}  presented constraints on the contribution of a non-thermal component to the total pressure which is assumed to be in hydrostatic equilibrium in the gravitational potential of the X-COP clusters.
The non-thermal pressure is parametrized as a power-law dependence upon the gas density with the normalization and the slope as free parameters.
These free parameters are constrained by requiring that the gas mass fraction at an overdensity of 500 and 200 $\rhocrit$ matches a ``universal" value, which was derived from cosmological simulations, accounting for the baryonic mass fraction and subtracting a statistical contribution for stellar mass \citep[see also][]{Eckert19}.
By construction, the model is sensitive to the regions where a gas mass fraction measurement (based on the initial assumption that the total pressure support is only thermal) is available, and it  can only account for non-thermal pressure support in systems where the observed gas fraction exceeds the universal value. 

For the sake of simplicity, we compare our $\alphaturb=0$ rotation velocity profile, which contains the entire information on the non-thermal support, with the 1D turbulent velocity dispersion profiles obtained in \citetalias{Ettori22} for the same cluster (see their Eq.~21).
In the regions where the gas mass fractions are available ($r \sim r_{500} - r_{200}$), the 1D turbulent velocity dispersion profiles of \citetalias{Ettori22} for most clusters 
lies within the range between the 16th and 84th percentiles of our reconstructed rotation velocity profiles.
However, in A1795 the 1D turbulent velocity dispersion profile from \citetalias{Ettori22} is significantly lower than the 16th percentile of our rotation velocity profile.
Although a detailed investigation of the cause of this discrepancy is beyond the scope of this comparison, we suggest a potential inconsistency in the mass estimates\footnote{The inferred $\Mtwo$ of our $\alphaturb = 0$ model for A1795, (1.37 $\pm$ 0.27) $\times 10^{15} \Msol$, differ from that of \citetalias{Ettori22}, 0.71 $\times 10^{15} \Msol$ (as inferred from the hydrostatic bias reported in their paper), by $2.5\sigma$. We verified that, assuming  0.71 $\times 10^{15} \Msol$ as median of the prior distribution of $\Mtwo$, we find a profile of $u_\phi$ consistent with the $\sigmaturb$ profile of \citetalias{Ettori22}.}
rather than issues related to the different modeling.
Overall, in \citetalias{Ettori22}, A2255 and A2142 show a central 1D velocity dispersion of about 400 $\kms$, and are therefore promising candidates for a detectable ICM rotation. 
While A2142 was found with low non-thermal support in the central regions in our model (see the left panel of Fig.~\ref{fig.sample_rot}), we confirm the high level of non-thermal support near the center of A2255.


In \citetalias{Bartalesi24}, we compared the rotation curves of our models with those recovered from some cosmologically simulated clusters. 
Given that the inferred constraint on the turbulence in our models impacts the inferred rotation, we investigate here how our constraints on the turbulent component of the ICM velocity field relate to the predictions of \citetalias{angelinelli20}, who accurately recovered the turbulent part of the ICM velocity field in their simulated sample (see Sect.~\ref{sec:intrinsic}, for details). 
In the lower-left panel of Fig.~\ref{fig.sample_turb}, we overlay the median profile of $\alphaturb$ from \citetalias{angelinelli20}.
This average profile is well in agreement with our median results, lying within the interval between the 16th and 84th percentiles of the distribution of the $\alphaturb$ profiles in the X-COP population from $\sim$ 50 kpc out to $\sim 3\rfive$, although local tensions in specific cases are expected and observed (see plots in Appendix~\ref{sec:profiles}).




\subsection{Detecting the ICM rotation with \textit{XRISM}: Abell~2255}
\label{sec:A2255}

In this section, we study the capability of X-ray spectroscopy at high resolution ($\sim$ eV) to detect and characterize the rotation of the ICM.
In particular, the X-ray calorimeter {\it Resolve} onboard the \textit{XRISM} mission\footnote{\url{https://www.xrism.jaxa.jp/en/}} is the only currently available instrument that can offer such a spectral resolution, although with a limited collecting area and Field of View (FoV) that make efficient only exposures of the brightest regions, i.e. the cores of massive nearby objects.
In \citetalias{Bartalesi24}, we studied the significance of the measurement of the rotation through the estimates of the centroid of the iron emission lines, the impact of the line broadening on these estimates and its cross-correlations with the other fitting parameters in the analyses of a set of \textit{XRISM/Resolve} mock spectra generated from three realistic (but generic) rotating ICM models. Here, we present a feasibility study applied to A2255, the most interesting target because of its highest median rotation velocity ($\approx 500 \, \kms$ as inferred from the $\alphaturb \neq 0$ model) near the center in the considered sample of objects.

A2255 is a disturbed, non-cool-core system at redshift $z_0 = 0.0809$, settled in the North Ecliptic Pole SuperCluster. 
It shows a flat X-ray surface-brightness profile in its core and an X-ray morphology extended along the E--W axis \citep{Sakelliou06}, which could indicate the presence of rotation.
Spectral analysis of {\it XMM-Newton} data shows that the temperature distribution presents azimuthal variations.
These characteristics were interpreted as due to a shock moving along the E--W direction with Mach number $\sim 1.24$ and with a core crossing occurred $\sim 0.15$ Gyr ago \citep{Sakelliou06}.
Also some properties inferred from multiwavelength observations, such as the fact that the X-ray centroid of the cluster does not coincide with the position of the brightest galaxy, are consistent with the merger scenario \citep{Burns95}.
While there is currently no solid evidence for a radio counterpart of E--W shock, many relics were detected in the outskirts (some seen in projection), and a radio halo was detected in the center \citep{Pizzo08, Pizzo09}.
Recent deep radio observations of A2255 suggested that very composite mechanisms, including seed particles injected by radio galaxies and spread in the ICM by turbulent motions and weak shocks, are at work at the same time to generate the radio halo \citep{Botteon20,Botteon22}.
Interestingly, \cite{Roettiger00}, using a hydrodynamical N-body simulation, found that off-axis mergers are an efficient mechanism for transferring angular momentum to the ICM, which suggests that large-scale rotation of the ICM could be present in post-merging objects \citep[see also][]{Roettiger98}.

\begin{figure}
   \centering
   \includegraphics[width=0.5\textwidth]{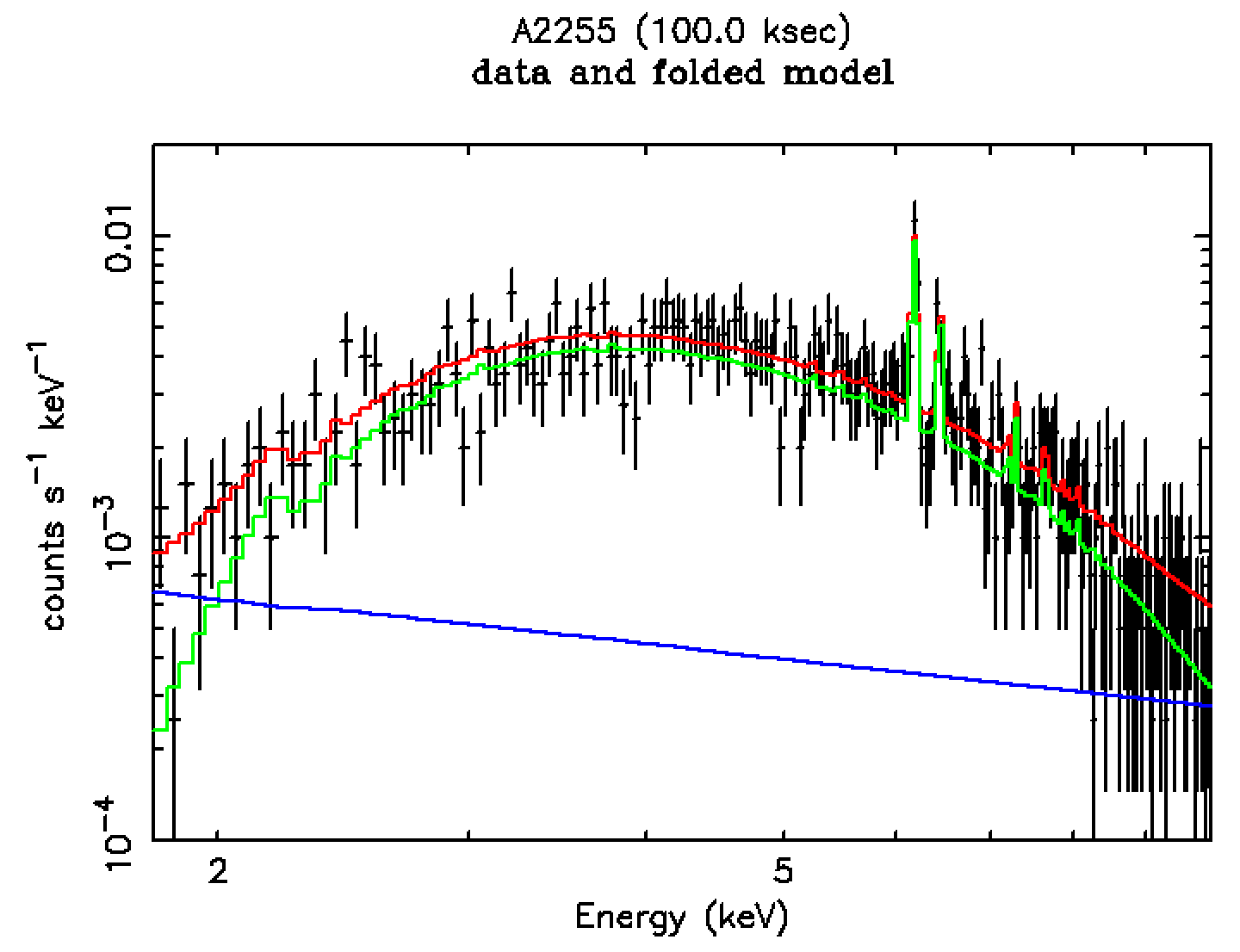}
   \caption{Simulated spectrum (black crosses), with an exposure of 100 k-seconds, in half of the {\it Resolve} FoV with the best-fit {\tt bapec} model overplotted. 
   Green line: thermal model only; blue line: instrumental background; red line: thermal model plus instrumental and cosmic background.
   The most prominent lines are the Fe XXV and Fe XXVI complexes in the range 6 - 7 keV. 
   }
    \label{fig.feasi}
\end{figure}

We simulate a pointing with \textit{XRISM/Resolve} centered on the core of A2255 and including a rotation speed of 500 $\kms$. 
When we extract a spectrum from the entire FoV of \textit{Resolve}, consisting of $6 \times 6$ pixels covering an area $\approx 3 \times 3$ arcmin$^2$, and affected by a PSF with a Half-Power Diameter of about 1.3 arcmin ($\approx$ 130 kpc, at the redshift of A2255), both turbulent and rotational motions would contribute to the measured broadening of the X-ray emission lines.
To associate the rotation with the shifting of the centroids of the X-ray emitting lines and decouple it from the turbulent motions in the ICM, we consider the following observational strategy.
We assume that, if the ICM rotation is present in the {\it Resolve} exposure, it can be resolved by looking at the different redshift measured in two complementary spectra extracted from the single observation.
To evaluate this, we divide the FoV into two complementary sectors of $3 \times 6$ pixels.  
We simulate the ICM emission using the thermal plasma emission model {\tt bapec}\footnote{ \url{https://heasarc.gsfc.nasa.gov/docs/xanadu/xspec/manual/XSmodelApec.html}} with a 6.2 keV temperature, the value spectroscopically measured with an XMM exposure of the central regions \citepalias[see][]{Ghirardini19} and abundance equal to 0.3 times the solar abundance in \citealt{A89} \citep[see][for details]{Ghizzardi21}. 
The normalization is set to the fraction ($\sim$ 20\%) of the 0.5 – 2 keV band luminosity estimated within the hydrostatic $\rfive$ that falls within the {\it Resolve} FoV using a $\beta$ model \citep{Cavaliere78} with parameters the scale radius $r_{\rm c} = 0.15 \rfive$ and the exponent $\beta =0.65$.
Given that we are considering half of the {\it Resolve} FoV, the normalization of the {\tt bapec} model is defined as half of the estimated flux. 
In the simulated spectrum, the nominal redshift $z_0$ is modified as $z_\mathrm{u} = (1 + z_0) \sqrt{ (1 + \upeak/c) / (1-\upeak/c) } -1$ \citep{Roncarelli18}, where $c$ is the light speed.
Given that the peak of the rotation lies within the $3\times 6$ sector, in this equation we considered the peak rotation velocity, $\upeak$ (positive for a receding ICM). 
For $z_0=0.0809$ and $\upeak \sim \pm500 \, \kms$, $z_\mathrm{u} = 0.0791$ and $0.0827$, respectively.
The most prominent emission line complex, key to accurately recovering the redshift of the ICM, is that of iron (Fe XXVI) at rest-frame energy equal to 6.7 keV.
The centroid of it should be shifted, in the observer's frame, to (6.188, 6.199, 6.209) keV for a rotation of $+500, 0, -500$ km/s, respectively.
These values correspond to differences in the locations of the Fe XXVI lines of $\Delta E \approx 10-21$ eV that cannot be resolved with CCD-like spectral resolution (like for XMM), but are well above the {\it Resolve} spectral resolution and accuracy of about 4.8 and 0.2 eV, respectively. 
We include a broadening of the X-ray emission lines equal to the median turbulent velocity dispersion ($\approx 300\, \kms$) inferred from our model.
All the spectra are absorbed by the expected Galactic column density of $2.5 \times 10^{20}$ cm$^{-2}$. 
This parameter is kept fixed in the following spectral analysis.
Using the responses for the closed Gate-Valve configuration, and adding an instrumental and a cosmic background and a Poissonian noise, we simulate in {\tt XSPEC} the 1.8 - 10 keV spectrum of the rotating ICM, as shown in Fig.~\ref{fig.feasi}.
Importantly, the instrumental background is significantly lower than the thermal model, and the cosmic background adds a relatively small contribution to the flux of the 6.1 - 6.2 keV emission line complex.
We proceed by fitting the broad-band (1.8 - 10 keV) spectrum, fixing the best-fit value of the temperature, and fitting the region around the Fe XXV-XXVI complex again. 
We obtain that a minimum exposure of 100 ks is required to reach accuracy and precision of $<1$\% on the measurement of the redshift, allowing us to identify the centroid of the redshifted emission lines at more than $5 \sigma$ level of confidence for a rotation greater than $\pm 400\, \kms$ ($\Delta E \approx 17$ eV), and at $>3\sigma$ for a rotation down to $\pm 200$ km/s ($\Delta E \approx 8.2$ eV).
The temperature, line broadening and metallicity are consistently measured with an accuracy in the order of 10 - 20\%. The PSF is not an issue when determining the centroid of the emitting lines in two different sectors. Moreover, based on the spectral simulations and their analyses presented in \citetalias{Bartalesi24}, no significant cross-correlation between the {\tt bapec} parameters affects our reconstruction, ensuring us on the possibility to prove the expected evidence for a rotating ICM in A2255.

\section{Conclusions}
\label{sec:conclusion}

In this work, we presented equilibrium models with rotating ICM for 11 galaxy clusters of the X-COP sample.
In the models, based on those of \citetalias{Bartalesi24}, the gas has axisymmetric composite polytropic distribution, in equilibrium in the gravitational potential of a spherical NFW dark-matter halo. For each cluster we considered both the case of turbulent ICM ($\alphaturb\geq 0$) and the case of ICM with negligible turbulence ($\alphaturb=0$). 

The profile of rotation velocity and the distribution of turbulent velocity dispersion are described with flexible functional forms, able to reproduce the radial profiles obtained from the spherically-averaged analyses of clusters formed in cosmological simulations.

We considered the observed profiles of the normalization of the thermal continuum (which scales with the gas density squared), of the spectroscopic gas temperature, and of the SZ-derived pressure as obtained by \citetalias{Ghirardini19} using XMM-\textit{Newton} and \textit{Planck} measurements.
We constrained the cluster gravitational potentials using estimates of the virial mass inferred either from the WL analysis by \cite{Herbonnet20} or under the assumption of a "universal" value of the baryon fraction by \cite{Eckert19}. 
For each cluster in our sample, we used an MCMC method to infer the posterior distributions of the parameters of the models using observational data on the thermodynamic profiles and the aforementioned constraints on the gravitational potential.
In the comparison with the observational datasets, we assume to observe our cluster models with line of sight orthogonal to the rotation and symmetry axis.

Rotation of the ICM has not been detected so far, but rotation-induced shifts of the centroids of X-ray emission lines could be measured with the currently operating \textit{XRISM/Resolve} X-ray spectrometer. 
To assess the feasibility of such rotation-velocity measurements, for our rotating-ICM model of the cluster Abell 2255, we computed and analyzed a mock spectrum of a central exposure with \textit{XRISM/Resolve}.


Our main findings are the following:
\begin{itemize}
    \item Our models reproduce well the observed radial profiles of the thermodynamic quantities of the ICM with median residuals of $\approx$ 5\%.
    \item According to our $\alphaturb=0$ models, there is room for non-negligible rotation in the ICM of massive clusters. In our sample of clusters, the median rotation-velocity-to-velocity-dispersion ratio $u_\phi / \sigmagas$ increases from $\approx$ 0.1 at $R = 0.1 \rtwo$ to $\approx$ 0.4 at $R$ = $\rtwo$.
    Significant deviation from this trend is found in Abell~2255, with a median $u_\phi / \sigmagas$ of $\approx$ 0.5 ($u_\phi \approx 600\, \kms$) at $0.1 \rtwo$. 
    \item  The room for ICM rotation is significant also when the contribution of turbulence to the pressure support is accounted for. 
    On average, the $\alphaturb\geq 0$ models of the clusters of our sample have outward-increasing median $u_\phi / \sigmagas$ profiles going from $\approx$ 0.1 at $0.1 \rtwo$ to $\approx$ 0.3 at $\rtwo$. 
    A2255 has a median $u_\phi / \sigmagas$ of $\approx$ 0.4 ($u_\phi \approx 500 \, \kms$) at $0.1 \rtwo$, making it a very promising candidate for the presence of a detectable rotation in the ICM. 
    \item We showed that in A2255 a rotation velocity of 400 $\kms$ at a distance of 100 kpc from the center (predicted by our model) can be detected at more than $5\sigma$ level of confidence with a 100 ks central exposure with XRISM/\textit{Resolve}. 
\end{itemize}

The analysis presented in this paper has some limitations.
For instance, we neglected the effect of bulk motions, any ICM inflows and outflows, and the relative motion between the ICM and the dark-matter distributions, commonly known as sloshing.
Moreover, in the presence of sharp temperature variations along the LOS, which might be the case in mergers, the estimate of the temperature based on the spectroscopic-like temperature may be inaccurate. 
This can bias the reconstruction of the cluster mass profile in our model and then its rotation curve.
Moreover, in our model we are working under the assumption that the peak of the ICM emission coincides with the minimum of the cluster gravitational potential. Any offset between the two in real clusters can bias the mass profile reconstruction.

Nonetheless, as reported in Sect.~\ref{sec:data}, by construction the X-COP clusters lack evidence of large-scale morphological disturbances (for instance, induced by major mergers and coherent bulk motions) both in their X-ray emission maps and in the spectral analysis, and are the closest representation of a symmetric ICM in equilibrium with the underlying potential in the nearby Universe \citep[][]{Ghirardini19}.
This guarantees an accurate reconstruction of the gas density and temperature distribution in these systems through an azimuthally-averaged analysis, where we have also included a treatment of some possible residual systematic effects that were considered as an additive term in the error budget and radially constant. 
However, disturbances limited to local regions, on scales of few hundreds kpc or less, are observed: for instance, the centroid  of the X-ray emission in A2255 differs from the X-ray peak significantly \citep[the value of the centroid-shift is the second highest one, just below what is measured in the well known merging object A2319; see Table~2 in ][]{Eckert22b}.
This effect is partially mitigated by using large radial bins in our analysis, and by excluding the central regions where this offset should have the larger impact on the reconstructed mass profile.

Additional source of systematics to the work can come from the inaccuracy or non-universality of the ``universal value" of the baryon fraction \citep[see Sect.~5.1 of][for a discussion]{Eckert19}, which could bias the marginal priors on $\Mtwo$ and $c_{200}$ and, consequently, the inferred rotation curve of the ICM, though we conservatively added in quadrature the typical weak lensing error to that of these mass estimates.

In parallel to the rotation of the ICM, also the rotation of the cluster galactic component has been object of study.  
\cite{Hwang07} searched for candidates for the presence of the rotation in the galactic components of a sample of 56 clusters. 
They identified 12 clusters exhibiting observational characteristics (such as a rotation-velocity-to-velocity-dispersion-ratio between $0.5$ and $0.7$ and a significant velocity gradient across the entire plane of the sky), which a dynamically significant rotation would induce.
However, none of these clusters are included in the X-COP sample.
Stronger indications of the rotation of member galaxies has been found in the inner regions of the clusters Abell S1013 and MACS J1206.2-0847, and A2107, with rotation-velocity-to-velocity-dispersion ratio of 
$\approx$ 0.15 \citep{Ferrami23} and $\approx$ 0.6 \citep[][]{Song18}, respectively.
It is worth mentioning that the ICM can have a non-negligible rotational support even in the absence of dynamically unimportant rotational motions in the galactic component, as shown in the simulated cluster discussed in \cite{Roettiger00}, where the collisional ICM significantly rotates whereas the collisionless galactic and dark matter components do not. 
Robust determination of the angular momentum in the galactic 
and hot-gas components of galaxy clusters, from optical and X-ray spectroscopy, respectively, would provide further constraints on the models describing the formation and evolution of cosmic structures.

Improving spectral (and spatial) resolution in the X-ray band, as planned with the proposed European (\textit{NewAthena}\footnote{\url{https://www.cosmos.esa.int/web/athena}}) and Chinese ({\textit{HUBS}\footnote{\url{http://hubs.phys.tsinghua.edu.cn}}) 
missions, will open a new era in the understanding of the velocity field of the ICM \citep[see][]{Roncarelli18,clerc19,zhang24}. 
At the same time, more sophisticated models will be needed to reproduce, simultaneously, the radial profiles of the thermodynamic quantities, and the shifting and broadening of the emitting lines of the ICM as resolved spatially and spectroscopically in the X-ray bands.
These models will provide robust assessments of the contribution of the bulk motions to the energy budget of the ICM and of the underlying mass distribution of galaxy clusters, with important implications for our knowledge of the thermal balance of the ICM and of cosmological structure formation.

\begin{acknowledgements}
We thank the anonymous referee for very useful comments that helped improve the article.
We thank Dominique Eckert for helpful discussions on data analysis.
T.B. acknowledges funding from the European Union NextGenerationEU.
S.E. acknowledges the financial contribution from the contracts
Prin-MUR 2022 supported by Next Generation EU (n.20227RNLY3 {\it The concordance cosmological model: stress-tests with galaxy clusters}), ASI-INAF Athena 2019-27-HH.0,
``Attivit\`a di Studio per la comunit\`a scientifica di Astrofisica delle Alte Energie e Fisica Astroparticellare''
(Accordo Attuativo ASI-INAF n. 2017-14-H.0),
and from the European Union’s Horizon 2020 Programme under the AHEAD2020 project (grant agreement n. 871158).
\end{acknowledgements}

\bibliography{refs}{}
\bibliographystyle{aa}

\begin{appendix}

\section{Offset between X-ray and SZ data}
\label{sec:systematics}
We present here the method used to account for the possible systematic offset between X-ray and SZ data.
The joint fitting to X-ray and SZ data has been performed under the assumption of a nuisance parameter \citepalias[e.g.][]{Ghirardini19} $\eta = \PSZ / (\nee \kboltz T)$, where $\nee$ and $T$ are inferred from the fitting to $\NNN$ and $\Tsp$ data (see Sect.~\ref{sec:Xrays}).
For each cluster, we inferred the marginal posterior of $\eta$: the corresponding 16th-84th percentile intervals are shown in the lower panel of Fig.~\ref{fig.eta}. 
In the upper panel of Fig.~\ref{fig.eta}, we plot the distribution of $\eta$ in the entire sample\footnote{The pdf of the entire sample is computed as follows. 
    We defined a normal pdf individually for each cluster, with the median and standard deviation corresponding to the median and the average between the 16th and 84th percentiles of the $\eta$ distribution of each cluster, respectively.
    Then we summed these 11 individual pdfs and divided the resulting function by 11.} obtained for the $\alphaturb \geq 0$ models: we inferred a median value of 1.03 and a $1\sigma$ scatter\footnote{We evaluated the $1\sigma$ scatter on $\eta$ as the interval between the 16th and 84th percentiles of the distribution.} of 0.05.
The result for the $\alphaturb = 0$ model is similar, with a median value of 1.02 and a $1\sigma$ scatter of 0.06.
$\eta = 1$ is the expected value when the clusters have random orientations in the plane of the sky and random shapes of the 3D ICM distribution \citep[e.g.][]{Kozmanyan19}. The fact that in both models $\eta = 1$ lies in the interval between the 16th and 84th percentiles of the distribution of $\eta$ inferred in the entire sample is an indication of the goodness of our joint analysis.

\begin{figure}
   \centering
   \includegraphics[width=0.49\textwidth]{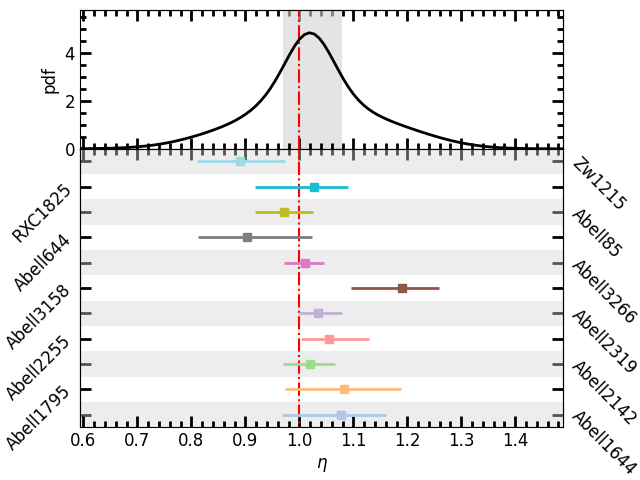}
    \caption{
    Upper panel. Probability density function (pdf) of the systematic offset parameter $\eta$ in the our cluster sample. The gray band ranges between the 16th and 84th percentile of $\eta$ in the entire sample.
    Bottom panel. Median $\eta$ of each cluster (as indicated in the vertical axes) according to the $\alphaturb \geq 0$, with error bars corresponding to the 16th and 84th percentiles. For reference $\eta = 1$ (no offset) is indicated with a red vertical line.
   }
    \label{fig.eta}
\end{figure}

\section{Set-up of the Markov Chain Monte Carlo method}
\label{sec:MCMC}

Using the \texttt{emcee}\footnote{\url{https://emcee.readthedocs.io/en/stable/}} sampler \citep{emcee}, we ran the Markov Chain and drew the samples of the parameter vector for each cluster model from the posterior $\Prob$.
Empirically, we noticed that a combination of \citet{Nelson13} and \citet{Goodman10} \texttt{moves}, as implemented in the \texttt{emcee} sampler, reduces the autocorrelation between successive iterations and optimizes the exploration of the parameter space.
We ran the MCMC for 60000 iterations with 500 chains. 
It is standard to consider the end of the burn-in phase when $\mathcal P$ is found to be stationary as the iteration number increases.
The burn-in phase is longer when $\alphaturb \ge 0$ than when $\alphaturb=0$, but as a rule it lasts less than 5000 steps. 
In few cases the number of steps of the burn-in phase is larger: $\approx 15000$ for both models of A2319.
To be conservative, in all cases we discarded the first 40000 steps.
We then applied a thinning of 700 steps to build the posterior sample.

\section{Flattening of the X-ray surface-brightness distributions}
\label{sec:flat}
Most X-ray surface brightness distributions of the X-COP clusters as resolved in the XMM-\textit{Newton} maps exhibit clear evidence for a deviation from circular symmetry.
Because of the presence of rotation, our models of the X-COP clusters  have oblate axisymmetric density distributions, even if the gravitational potential is assumed spherically symmetric.
As we assumed that the models are observed edge-on, their surface brightness distributions are flattened.
Here we compare the flattening of the models with that observed in the XMM maps.

\begin{figure}
   \centering
   \includegraphics[width=0.49\textwidth]{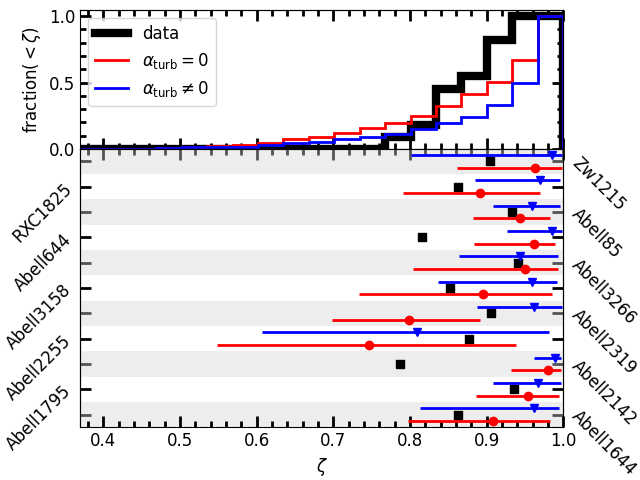}
    \caption{
    Observed and model axial ratio $\zeta$ of the X-ray surface-brightness distributions of the clusters of our sample.  
    Upper panel. Observed distribution of $\zeta$ (black) compared to the distributions of $\zeta$ of the $\alphaturb = 0$ (red) and $\alphaturb \geq 0$ (blue) models.
    Lower panel. Observed (squares) $\zeta$ compared with the median values of $\zeta$ obtained for each cluster (as indicated in the  vertical axis) for the   $\alphaturb = 0$ (circles) and $\alphaturb \geq 0$ (triangles) models. 
    The error bars indicate the 16th and 84th percentiles measured in our models. 
The observed data points have very small     $1\sigma$ uncertanties ($\delta \zeta / \zeta  \sim 10^{-3}$) not shown in the plot. 
    }
    \label{fig.axial_ratio}
\end{figure}

For the models we evaluated the surface-brightness distribution in each point of the plane of the sky $(y, z)$ as
\begin{equation}
\label{eq.brightness}
\Sigma(y,z)=2 \Lambda \int_{|y|}^{\rout} \frac{\niii(R, z) \nee(R, z) R \diff R}{\sqrt{R^2 - y^2}},
\end{equation}
where $\niii = n - \nee$
and $\Lambda $ is the cooling function, which we assume to be independent of temperature (and then of position), as appropriate when considering the soft X-ray band typically used for imaging.
We limited ourselves to the region in the plane of the sky, defined by 1 kpc $\leq |y| \leq \Rhatend$ and 1 kpc $\leq |z| \leq \Rhatend$, where $\Rhatend$ is  the outer radius of the $\Nbins$-th radial bin in the plane of the sky (see Sect.~\ref{sec:Xrays}).

Following \citetalias{Bartalesi24} (see their section~4.2), we use as measure the flattening of the X-ray 2D surface brightness distribution the axial ratio of isophotes, $\zeta$, computed from  the eigenvalues of the $2 \times 2$ matrix containing the centroids in $y$, $z$ and $yz$ weighted over the surface brightness. 
For each cluster model, we randomly extracted 100 parameter vectors from the posterior sample and computed $\zeta$ for each parameter vector.


In the upper panel of Fig.~\ref{fig.axial_ratio}, we compare the distribution of $\zeta$ in our models to those directly measured in the real clusters.
The $\zeta$ distribution of the $\alphaturb = 0$ models (with median 0.91) is more similar to the observed distribution of $\zeta$ (with median 0.90) than that of the $\alphaturb \ge 0$ models (with median 0.97). 
However, we note in the $\alphaturb = 0$ distribution a non-negligible tail at $\zeta < 0.75$ with no counterpart in the observed distribution.
The lower panel of Fig.~\ref{fig.axial_ratio}, where we compare the $\zeta$ measurements in the models individually with the observed $\zeta$ of each cluster, shows that the models of A1795 and A2319 perform better when  $\alphaturb \geq 0$ than when $\alphaturb = 0$, while for the other clusters the intervals between the 16th and 84th percentiles in the two models have substantial overlap.
In A644 and A2142, both $\alphaturb = 0$ and $\alphaturb \ge 0$ models do not reproduce the observed $\zeta$ at $1\sigma$ confidence level.
This might indicate that, in addition to the rotation of the ICM, other properties not accounted for by our models, such as deviations from spherical symmetry of the gravitational potential, non-negligibly contribute to the observed flattenings.



\section{Geometrical approximations in the model-data comparison}
\label{sec:approx}

Let us consider a generic quantity $Q(R, z)$ that depends on the intrinsic properties of the ICM in an axisymmetric cluster model represented in cylindrical coordinates $(R,\phi,z)$. In addition to the cylindrical coordinate system, we adopt also a Cartesian system with the same $z$ axis, and with $x$ and $y$ related to $R$ and $\phi$ by $R=\sqrt{x^2 + y^2}$ and $\phi = \arctan(x / y)$. 
Taking $x$ parallel to the line of sight, we also define a system of polar coordinates in the plane of the sky $(\Rhat,\phihat)$, related to $y$ and $z$ by $\Rhat = \sqrt{y^2 + z^2}$ and $\phihat = \arctan\left( z / y \right)$.
For comparison with the observations we need to evaluate integrals of $Q$ over a cylindrical shell with symmetry axis along the line of sight $x$, and with inner and outer radii $\Ri$ and $\Rplus$, respectively:
\begin{equation}
\label{eq:pqi}
P_{Q,i}=    
    \int_0^{2\pi} \int_{\Ri}^{\Rplus} f (\Rhat, \phihat)  \Rhat \diff \Rhat \diff \phihat,
\end{equation}
where $f(\Rhat, \phihat) \equiv 2 \int_0^{\infty} Q(R, z) \diff x$.
We also need to evaluate integrals over spherical shells with inner and outer radii $\radj$ and $\radplus$, respectively:
\begin{equation}
\label{eq:iqj}
I_{Q,j}=  \frac{\int_{\radj}^{\radplus} Q(R, z) r^2 \diff r}{\int_{\radj}^{\radplus} r^2 \diff r} .
\end{equation}
Given that our models are neither spherically symmetric nor circularly symmetric in the plane of the sky, the numerical calculation of Eqs.\ (\ref{eq:pqi}) and (\ref{eq:iqj}) 
is computationally expensive and would represent the bottleneck of our MCMC algorithm. 
We thus chose to replace Eqs.\ (\ref{eq:pqi}) and (\ref{eq:iqj}) with less time-consuming approximations. 


We limited the evaluation of $Q$ to $z=0$ plane. 
Assuming $f(\Rhat, \phihat) \approx f(\Rhat, 0)$ and $Q(R, z) \approx Q(R, 0)$, Eqs. (\ref{eq:pqi}) and (\ref{eq:iqj}) become
\begin{eqnarray}
\label{eq.appXraynew}
P_{Q,i}
&\approx&  2\pi \int_{\Ri}^{\Rplus} f (|y|, 0)  y \diff y
\end{eqnarray}
and 
\begin{eqnarray}
\label{eq.appSZ}
I_{Q,j}
&\approx& \frac{3\int_{\radj}^{\radplus} Q(|y|, 0) y^2 \diff y}{\radplus^3 - \radj^3},
\end{eqnarray}
respectively.

\section{Profiles and corner plots of individual clusters}
\label{sec:profiles}

We report here, for each cluster, profiles of some physical quantities and a corner plot for a selection of MCMC parameters.


\begin{figure*}
   \centering
   \hbox{
   \includegraphics[width=0.33\textwidth]{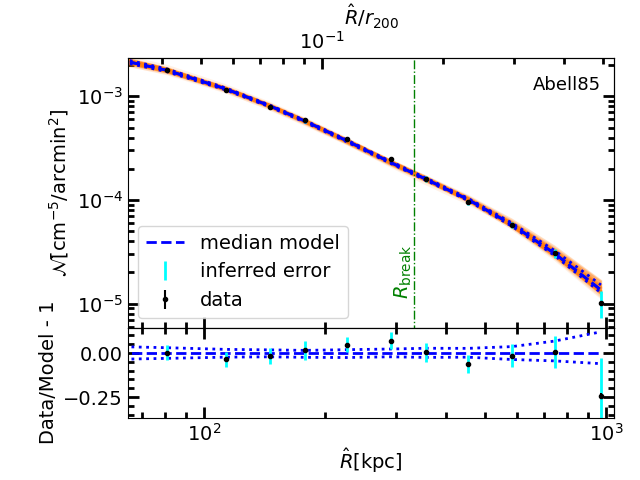}
   \includegraphics[width=0.33\textwidth]{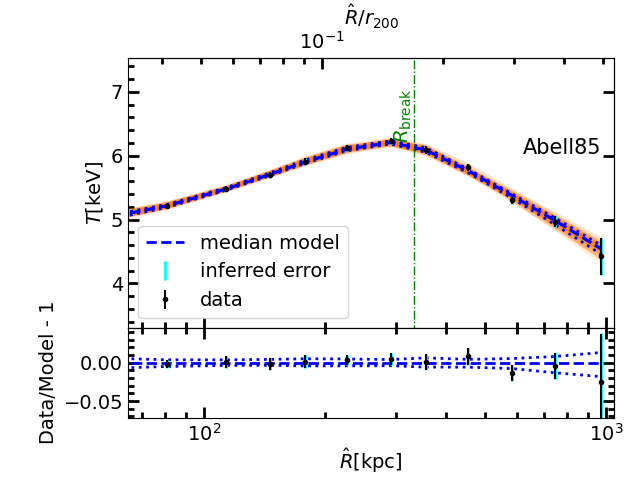}
   \includegraphics[width=0.33\textwidth]{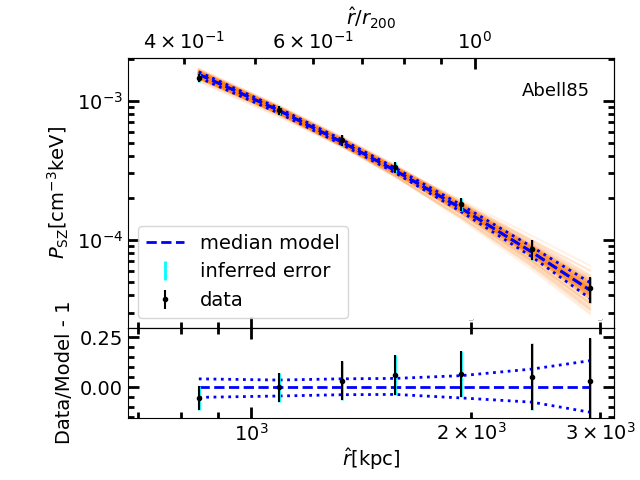}
   }
   \hbox{
   \includegraphics[width=0.33\textwidth]{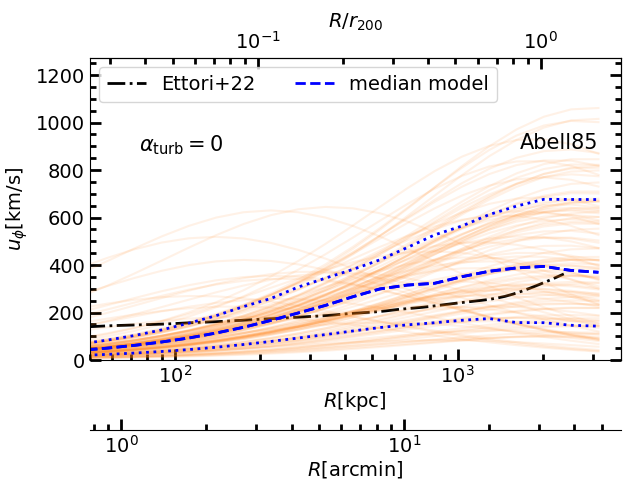}
   \includegraphics[width=0.33\textwidth]{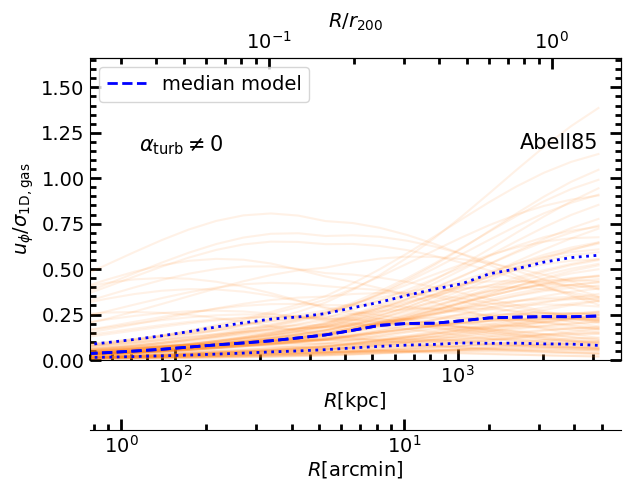}
   \includegraphics[width=0.33\textwidth]{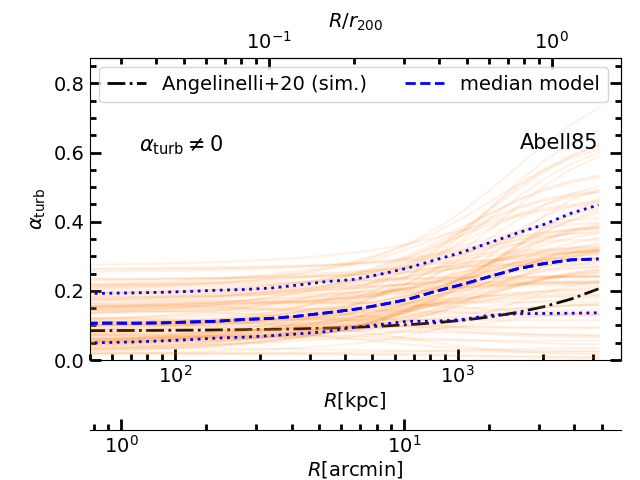}
} \includegraphics[width=0.49\textwidth]{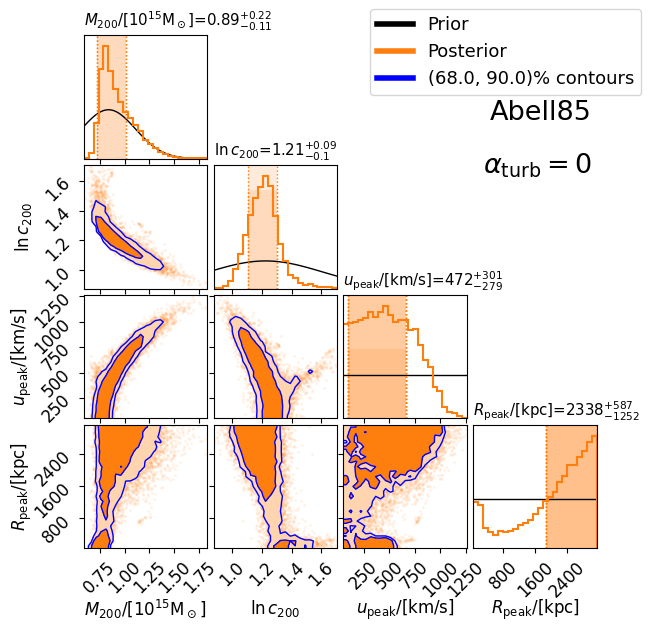}   
    \caption{Profiles and corner plot of A85. 
    {\em Top panels.} 
    Profiles of the normalization of the thermal continuum $\NNN$ (left panel), of the temperature $T$ (middle panel), and of the SZ pressure $\PSZ$ (right panel).  $\NNN$ and $T$ are plotted as functions of the projected radius $\Rhat$, while $\PSZ$ as a function of the intrinsic radius $r$.  In the upper part of each panel, the curves represent the inference on the profile of the $\alphaturb=0$ model: the dashed curve is the median profile, the dotted curves indicate the interval between the 16th and 84th percentiles, and the solid curves are 100 profiles randomly extracted from the sampling of the posterior; the points are the observational data, with the black error-bars indicating the statistical error, $\sigmastat$, and the cyan error-bars the inferred error, $\sqrt{1 + \epsilon^2} \sigmastat$ (see Sect.~\ref{sec:likelihood}, for details).  The lower part of each panel plots the residuals between the data and the median of the model shown in the corresponding upper part. 
        {\em Middle panels.} ICM rotation curve according to the $\alphaturb = 0$ model (left panel) and profiles of rotation-velocity-to-velocity-dispersion ratio (middle panel) and of turbulent-pressure-to-total-pressure ratio $\alphaturb$ (right panel), according to the $\alphaturb \geq 0$ model ($R$ is the radius in the equatorial plane).  The meaning of the dashed, dotted, and solid curves is the same as in the top panels.   The dot-dashed curve indicates the turbulent velocity dispersion profile estimated by \citetalias{Ettori22} in the left panel and the $\alphaturb$ profile of \citetalias{angelinelli20} (the same as in Fig.~\ref{fig.sample_turb}) in the right panel.
    {\em Bottom panel.}
    Marginal (diagonal panels) and two-parameters joint (off-diagonal panels) distributions of the MCMC parameters $\Mtwo$, $\ctwo$, $\upeak$ and $\Rpeak$ in the $\alphaturb = 0$ model (see Sect.~\ref{sec:intrinsic}).
    In each diagonal panel, the black curve and the orange histogram are the marginal prior and posterior, respectively;  
    the light-orange vertical band is the 68\% credible interval of the marginal posterior (the median and the 16th and 84th percentiles of the marginal posterior are reported over the top axis).
    In each off-diagonal panel, the dark-orange and light-orange regions, enclosed by the blue lines, define the 68\% and 90\% credible regions of the two-parameters joint posterior, respectively.}
    \label{fig.A85}
\end{figure*}

\begin{figure*}
   \centering
   \hbox{
   \includegraphics[width=0.33\textwidth]{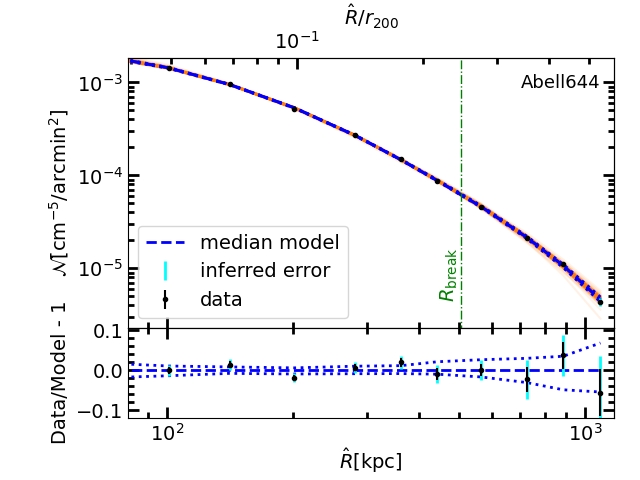}
   \includegraphics[width=0.33\textwidth]{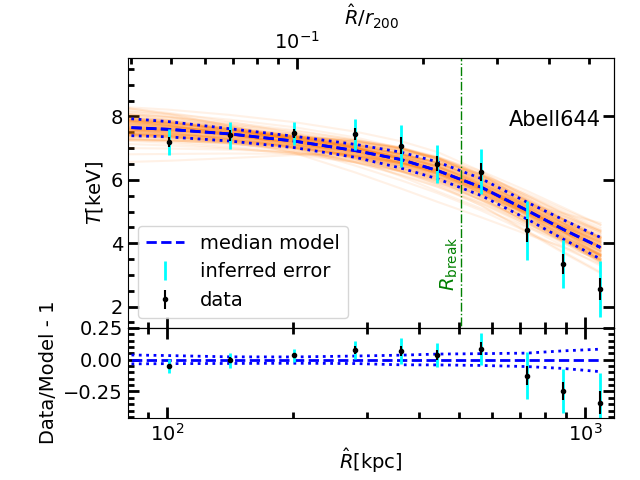}
   \includegraphics[width=0.33\textwidth]{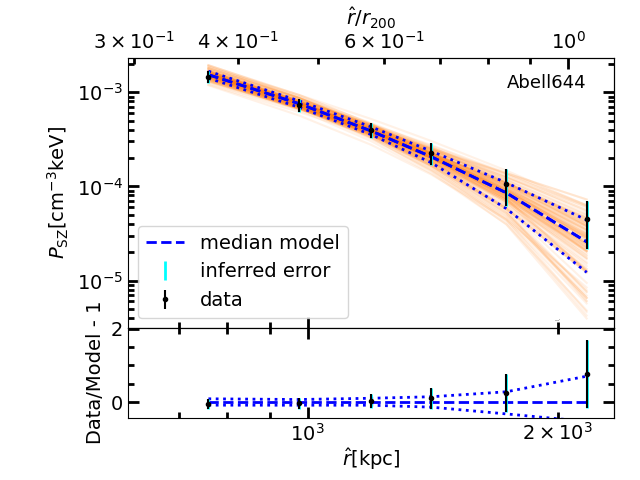}
   }
   \hbox{
   \includegraphics[width=0.33\textwidth]{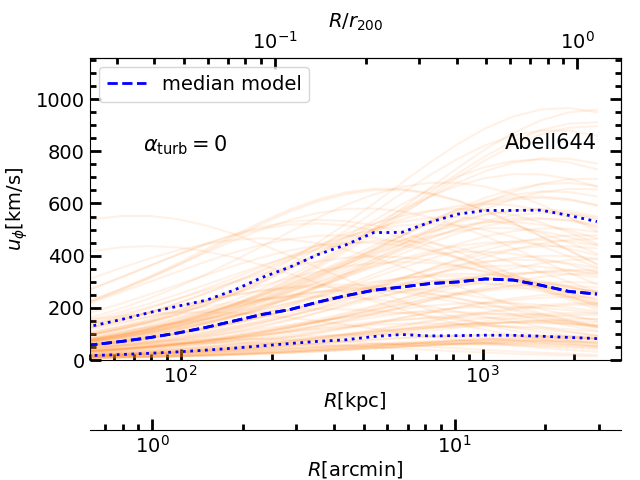}
   \includegraphics[width=0.33\textwidth]{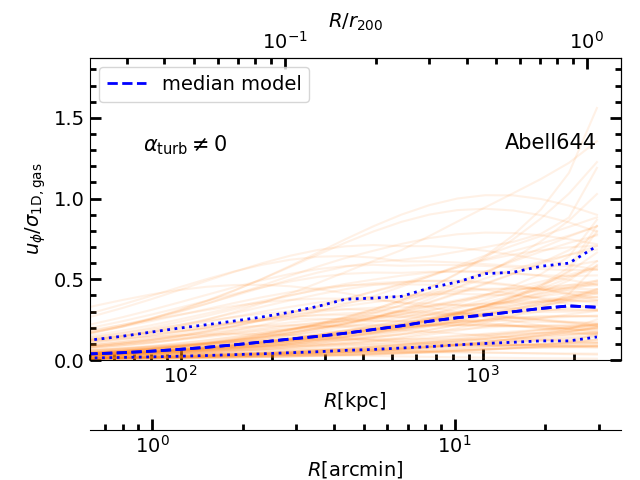}
   \includegraphics[width=0.33\textwidth]{Figures/turbulent_rotating_ICM_joint_X_SZ_datasets_Abell644_anisotropic_support.png}
} \includegraphics[width=0.49\textwidth]{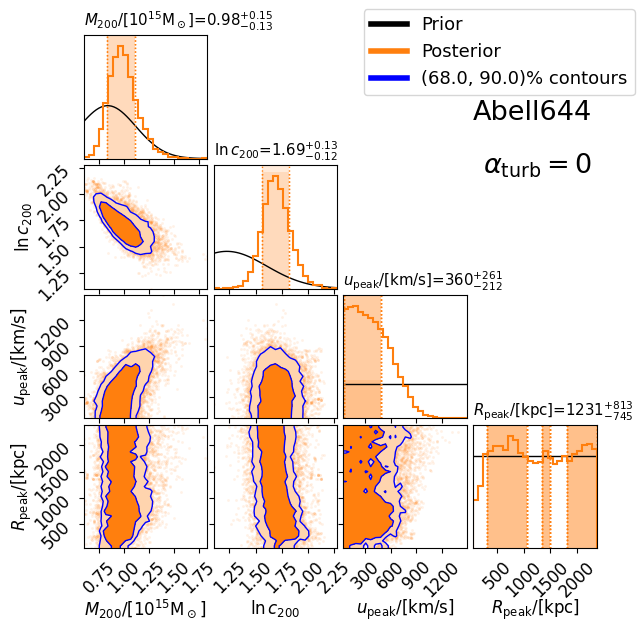}   
    \caption{Same as \ref{fig.A85}, but for A644.}
    \label{fig.A644}
\end{figure*}

\begin{figure*}
   \centering
   \hbox{
   \includegraphics[width=0.33\textwidth]{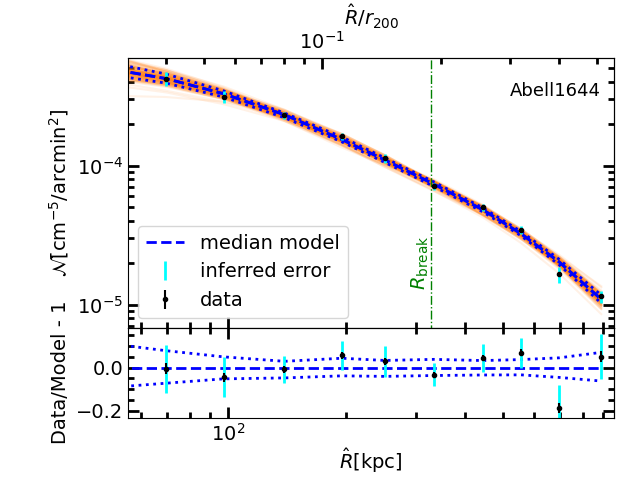}
   \includegraphics[width=0.33\textwidth]{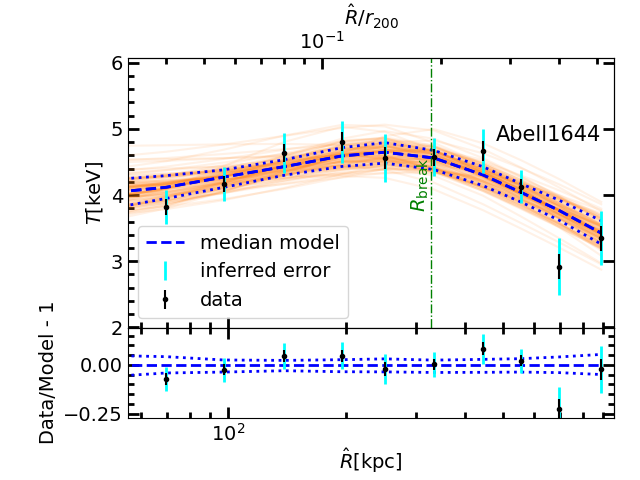}
   \includegraphics[width=0.33\textwidth]{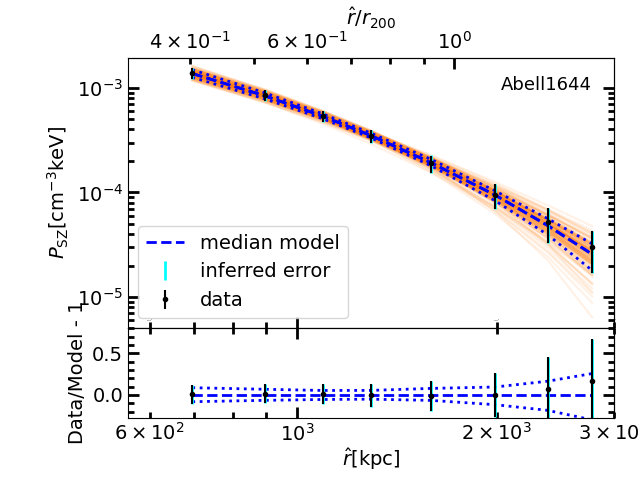}
   }
   \hbox{
   \includegraphics[width=0.33\textwidth]{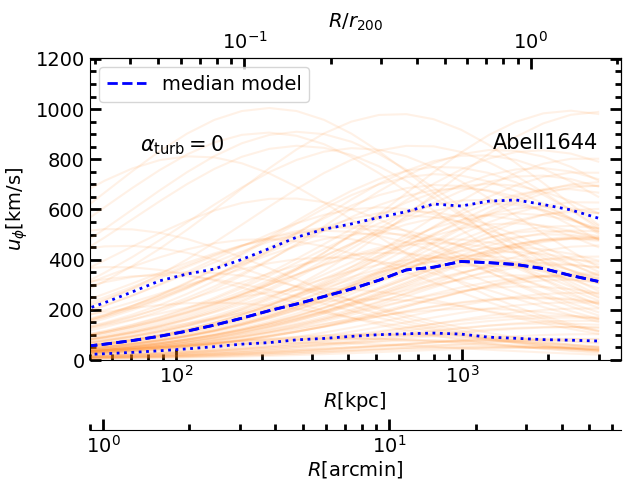}
   \includegraphics[width=0.33\textwidth]{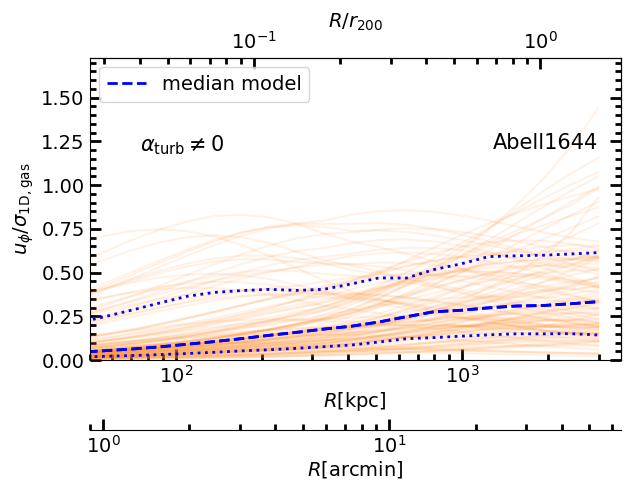}
   \includegraphics[width=0.33\textwidth]{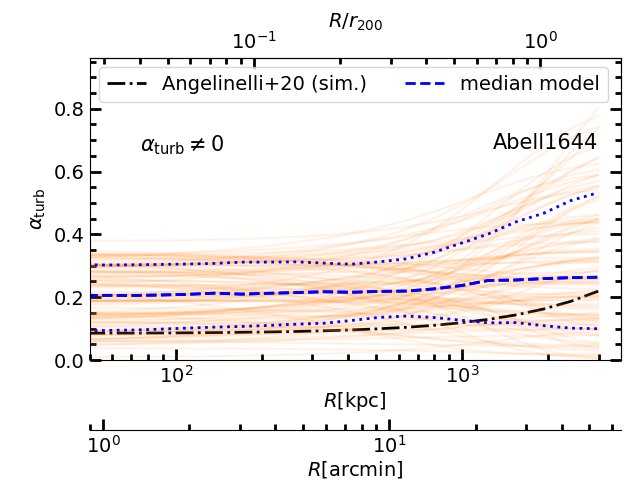}
} \includegraphics[width=0.49\textwidth]{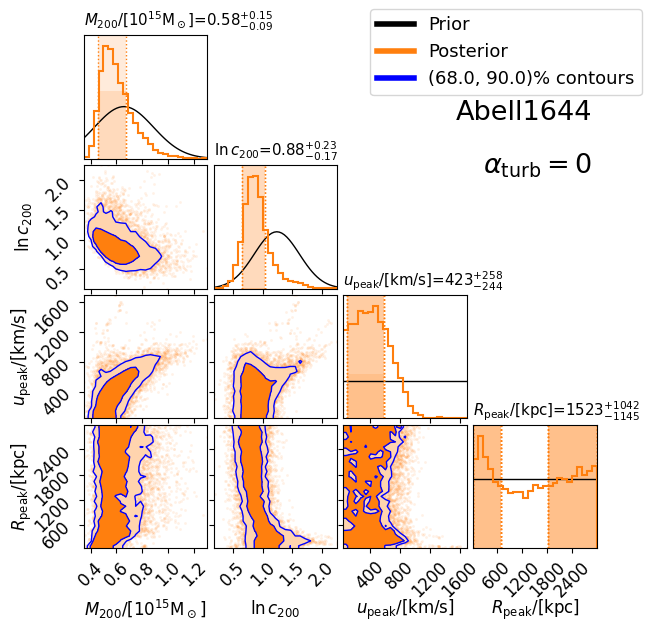}   
    \caption{Same as \ref{fig.A85}, but for A1644.}
    \label{fig.A1644}
\end{figure*}

\begin{figure*}
   \centering
   \hbox{
   \includegraphics[width=0.33\textwidth]{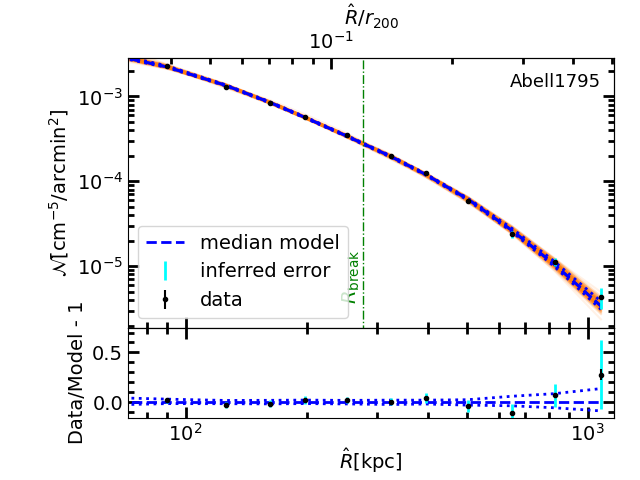}
   \includegraphics[width=0.33\textwidth]{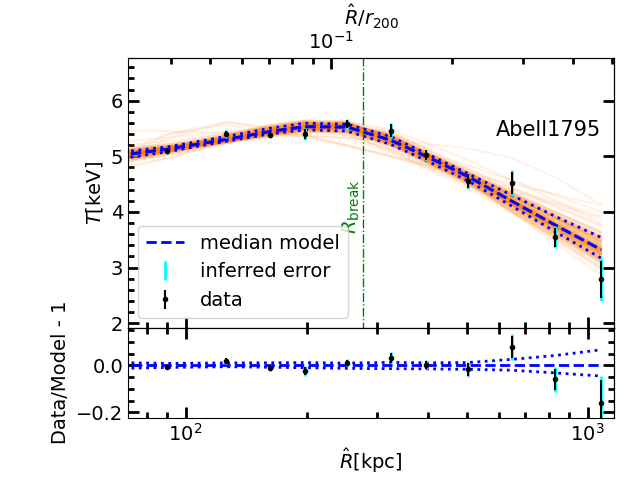}
   \includegraphics[width=0.33\textwidth]{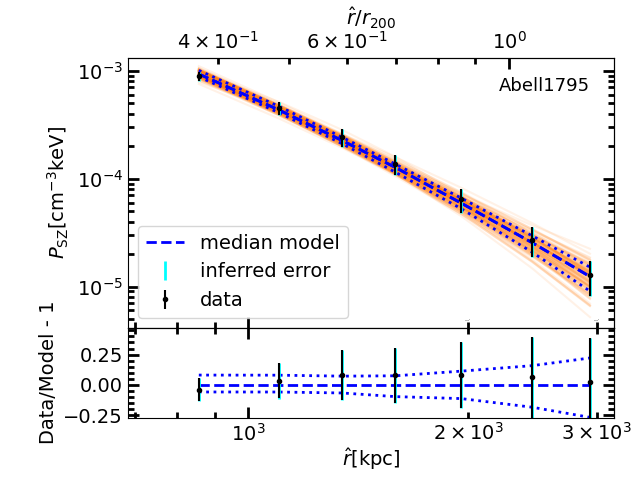}
   }
   \hbox{
   \includegraphics[width=0.33\textwidth]{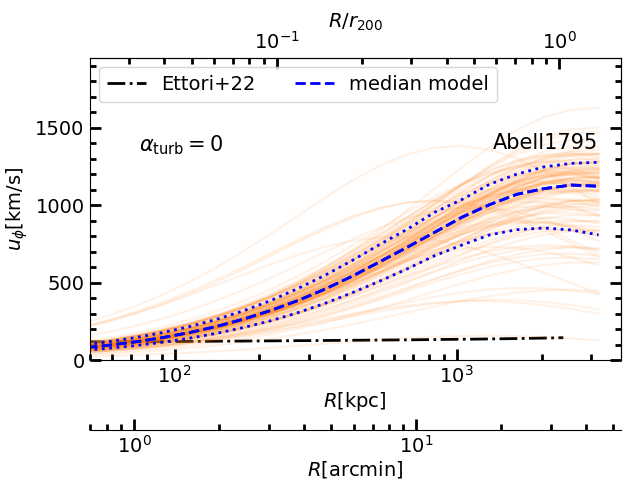}
   \includegraphics[width=0.33\textwidth]{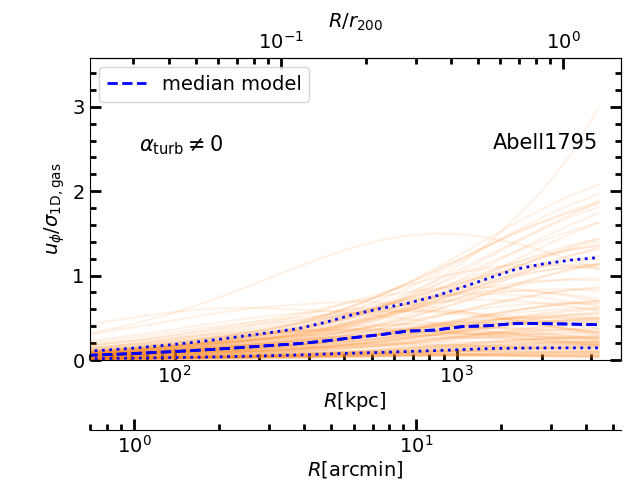}
   \includegraphics[width=0.33\textwidth]{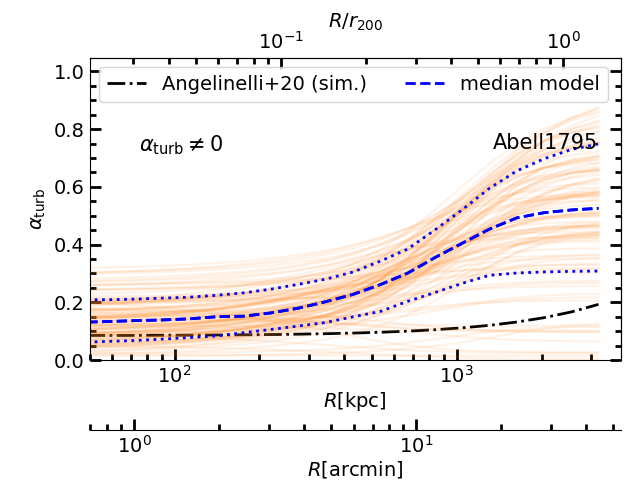}
} \includegraphics[width=0.49\textwidth]{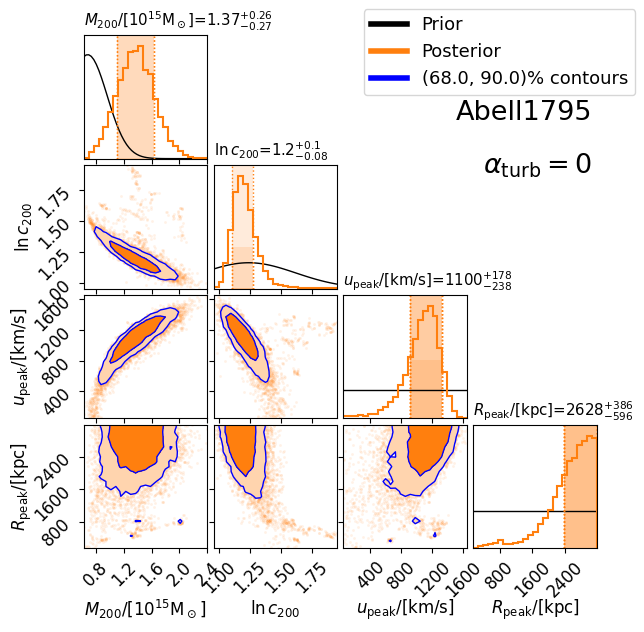}   
    \caption{Same as \ref{fig.A85}, but for A1795.}
    \label{fig.A1795}
\end{figure*}

\begin{figure*}
   \centering
   \hbox{
   \includegraphics[width=0.33\textwidth]{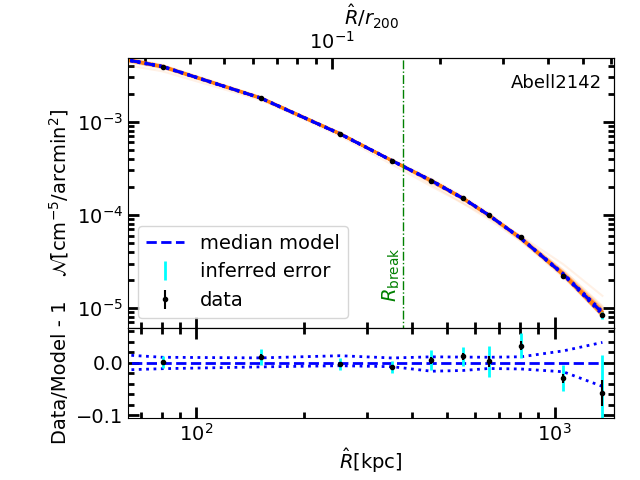}
   \includegraphics[width=0.33\textwidth]{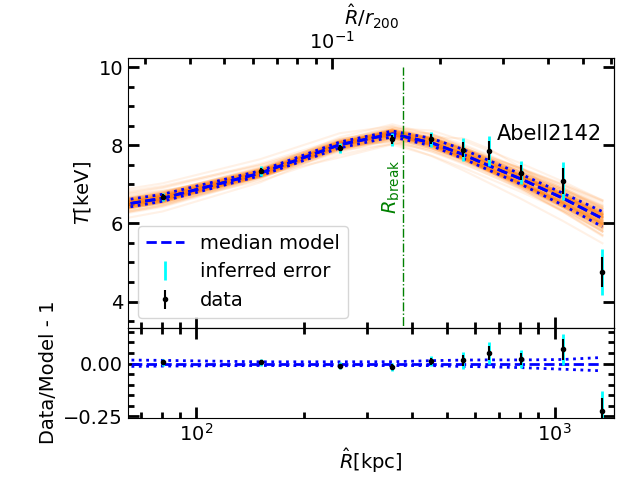}
   \includegraphics[width=0.33\textwidth]{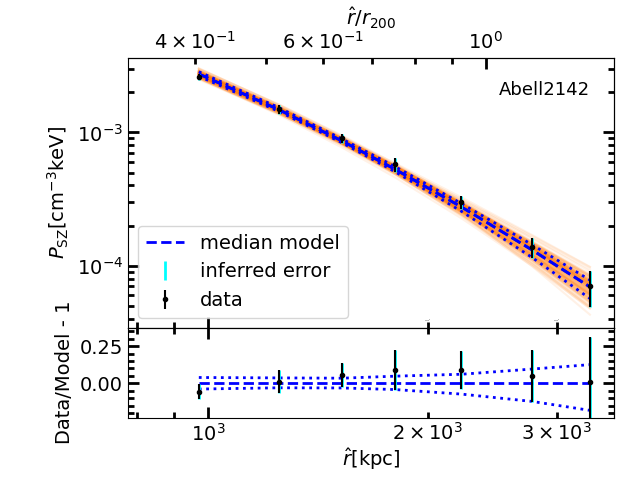}
   }
   \hbox{
   \includegraphics[width=0.33\textwidth]{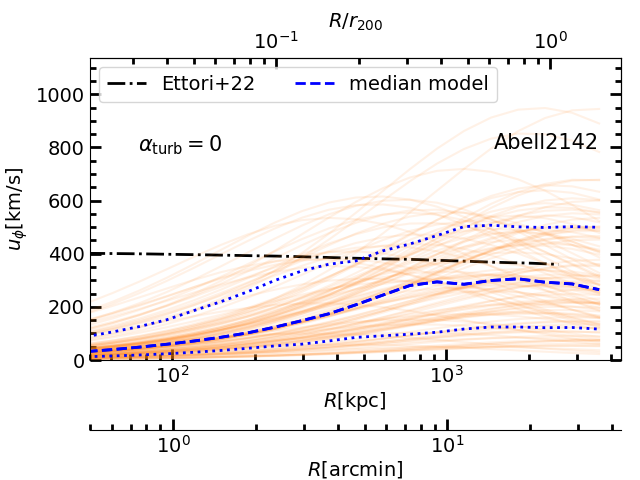}
   \includegraphics[width=0.33\textwidth]{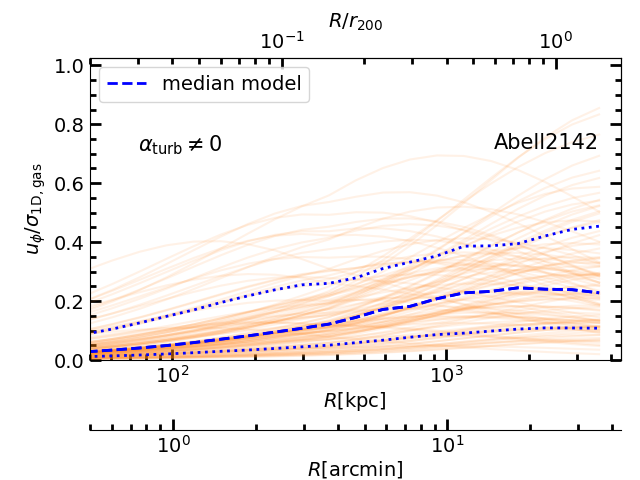}
   \includegraphics[width=0.33\textwidth]{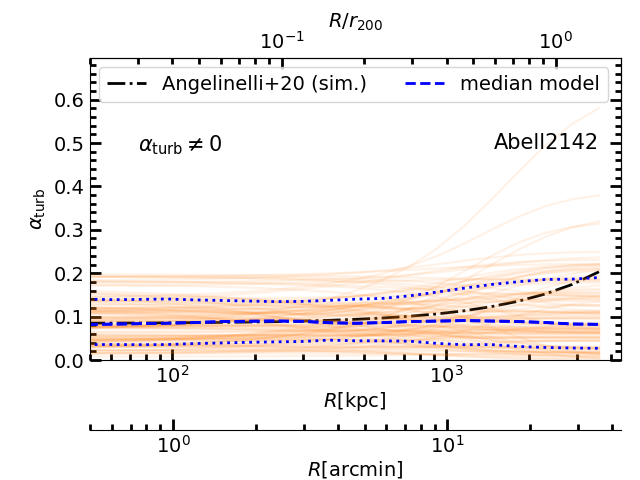}
} \includegraphics[width=0.49\textwidth]{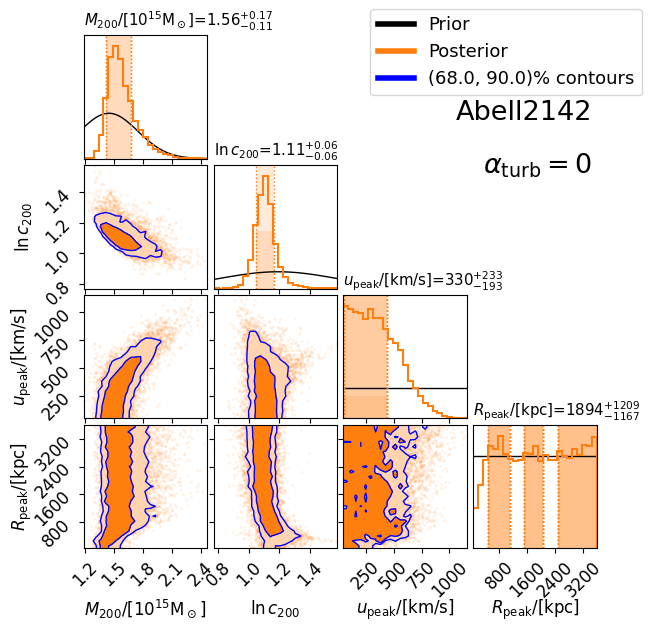}   
    \caption{Same as \ref{fig.A85}, but for A2142.}
    \label{fig.A2142}
\end{figure*}

\begin{figure*}
   \centering
   \hbox{
   \includegraphics[width=0.33\textwidth]{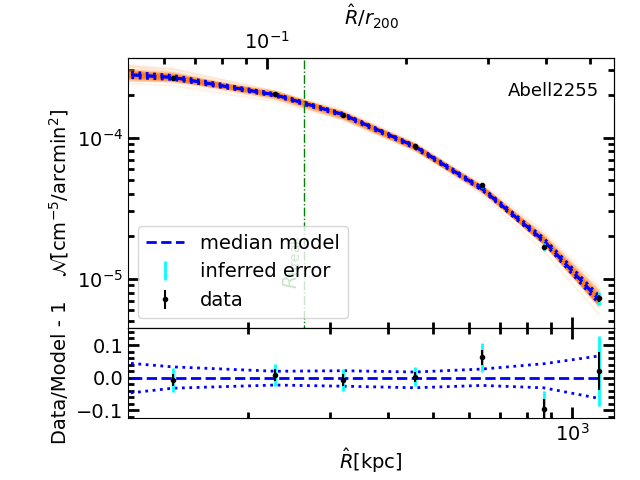}
   \includegraphics[width=0.33\textwidth]{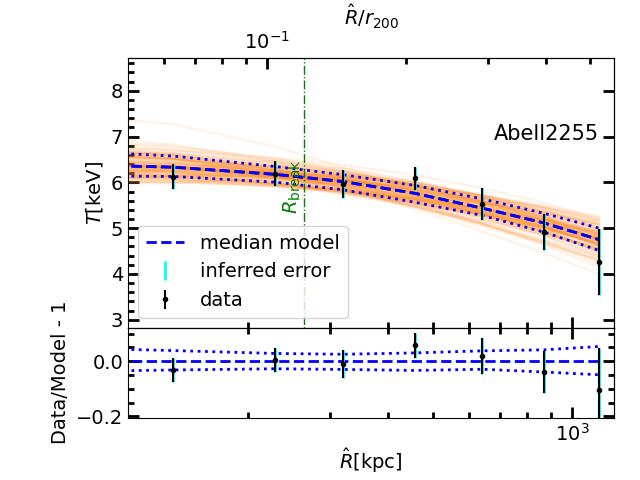}
   \includegraphics[width=0.33\textwidth]{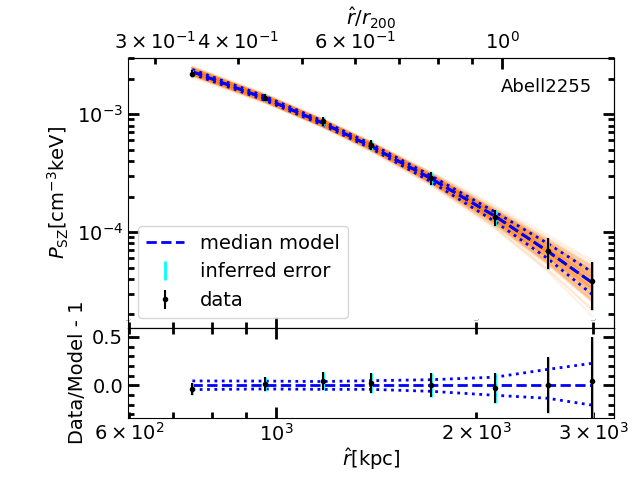}
   }
   \hbox{
   \includegraphics[width=0.33\textwidth]{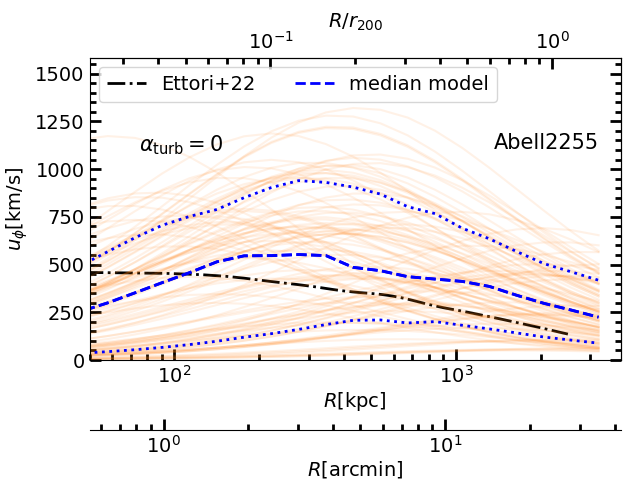}
   \includegraphics[width=0.33\textwidth]{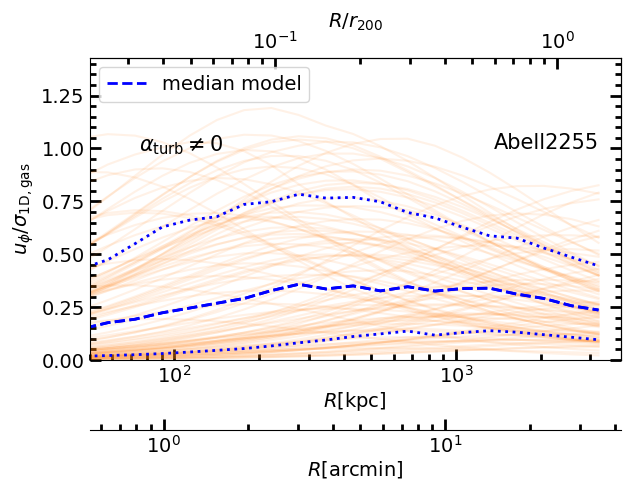}
   \includegraphics[width=0.33\textwidth]{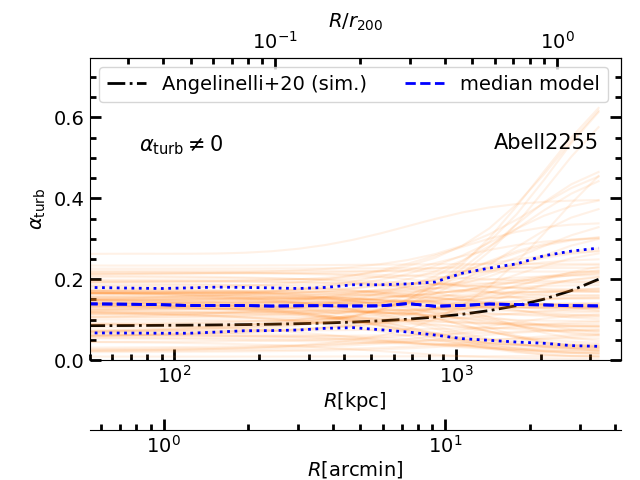}
} \includegraphics[width=0.49\textwidth]{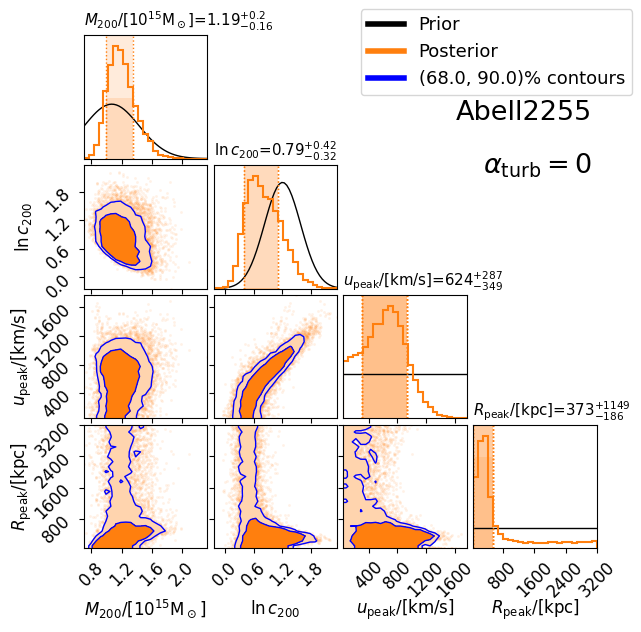}   
    \caption{Same as \ref{fig.A85}, but for A2255.}
    \label{fig.A2255}
\end{figure*}

\begin{figure*}
   \centering
   \hbox{
   \includegraphics[width=0.33\textwidth]{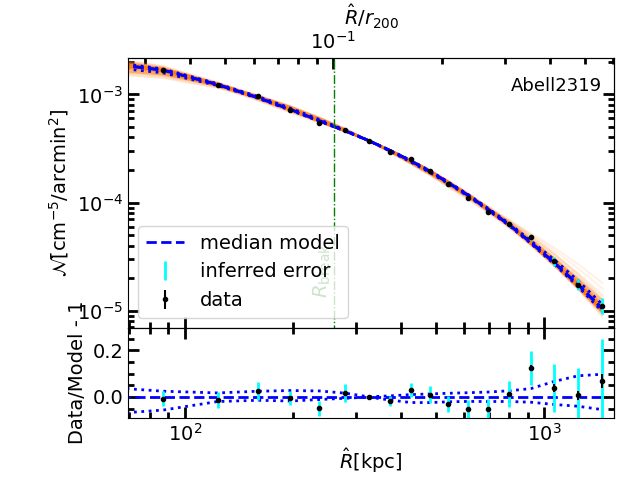}
   \includegraphics[width=0.33\textwidth]{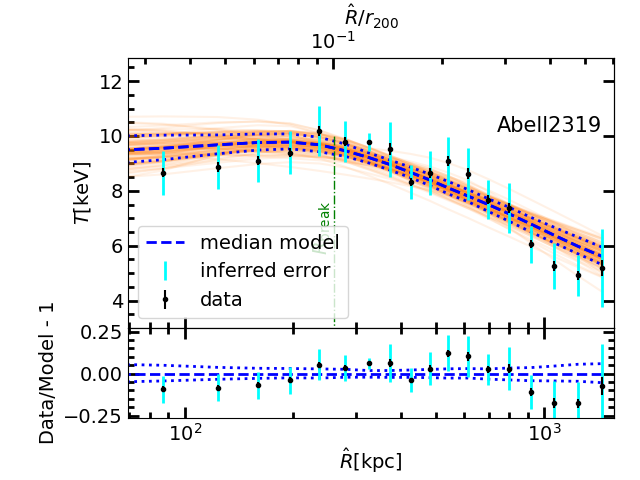}
   \includegraphics[width=0.33\textwidth]{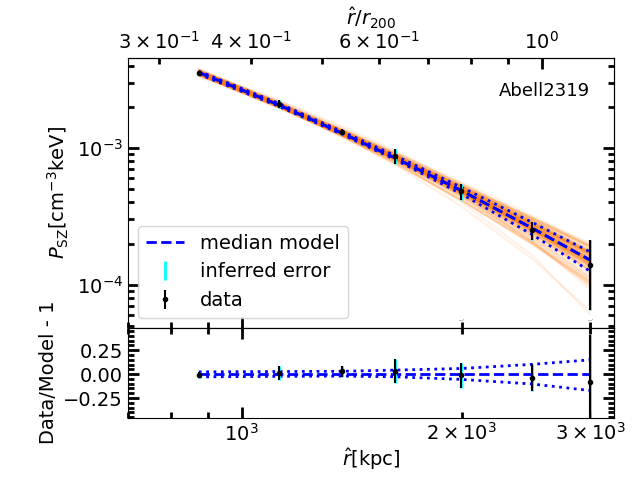}
   }
   \hbox{
   \includegraphics[width=0.33\textwidth]{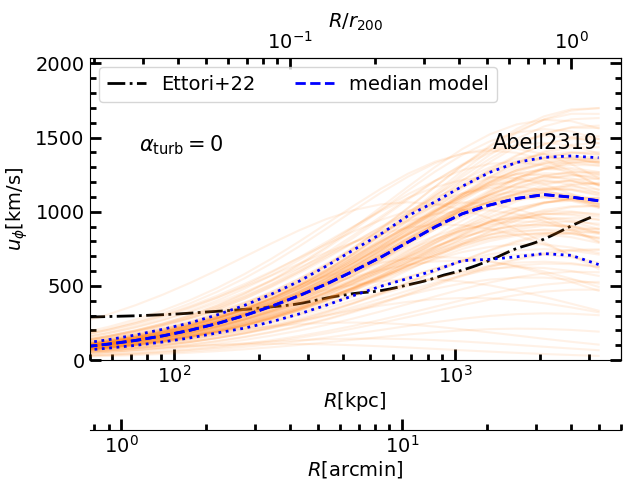}
   \includegraphics[width=0.33\textwidth]{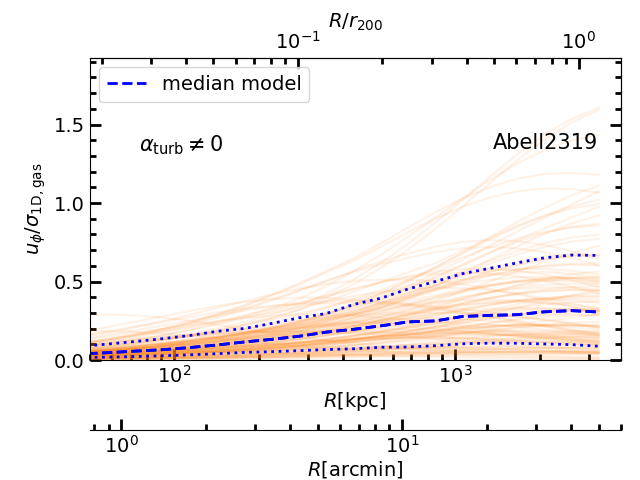}
   \includegraphics[width=0.33\textwidth]{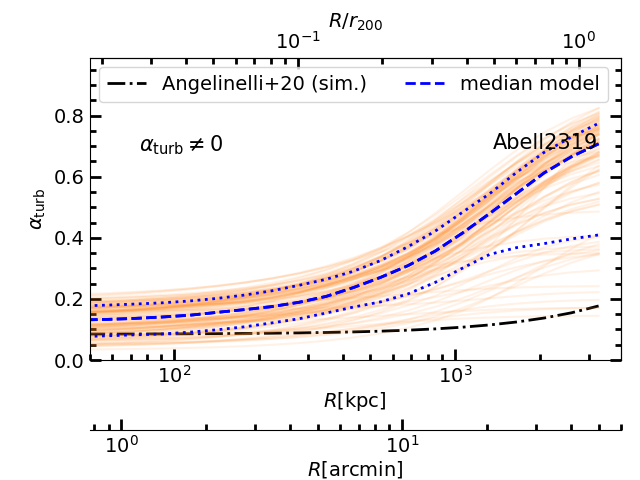}
} \includegraphics[width=0.49\textwidth]{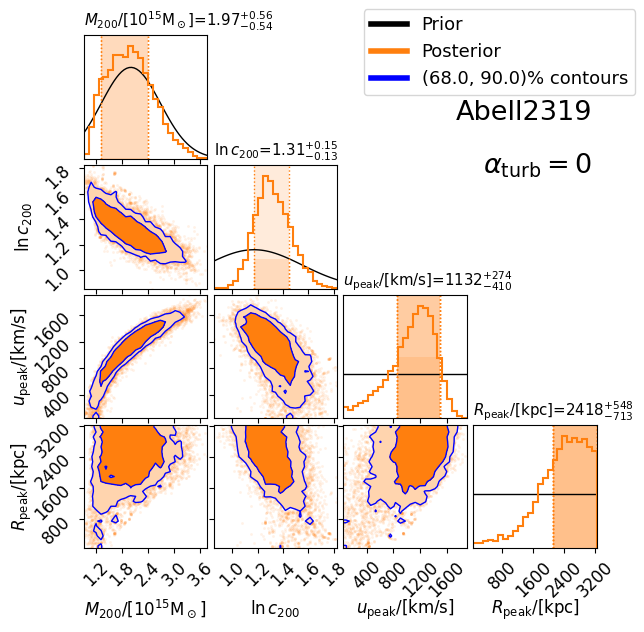}   
    \caption{Same as \ref{fig.A85}, but for A2319.}
    \label{fig.A2319}
\end{figure*}

\begin{figure*}
   \centering
   \hbox{
   \includegraphics[width=0.33\textwidth]{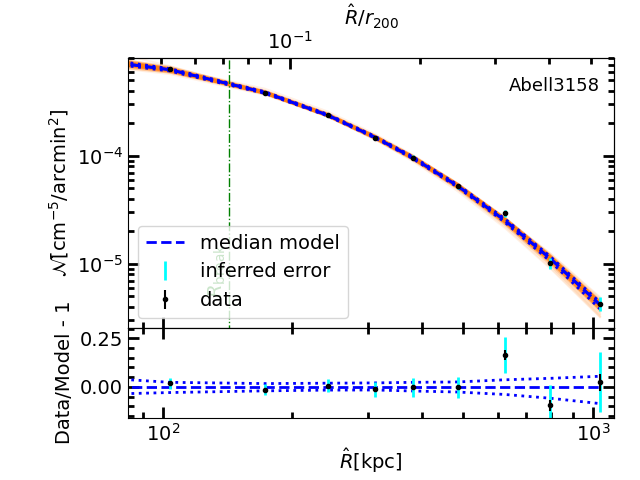}
   \includegraphics[width=0.33\textwidth]{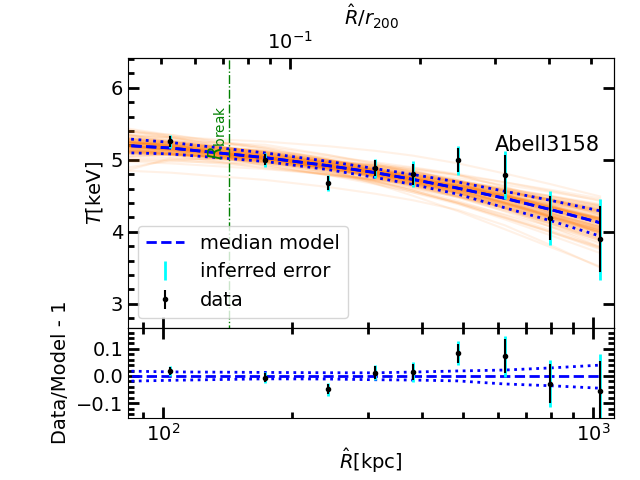}
   \includegraphics[width=0.33\textwidth]{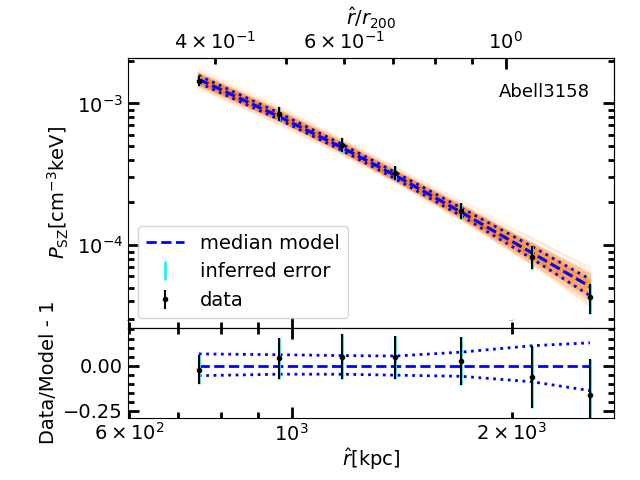}
   }
   \hbox{
   \includegraphics[width=0.33\textwidth]{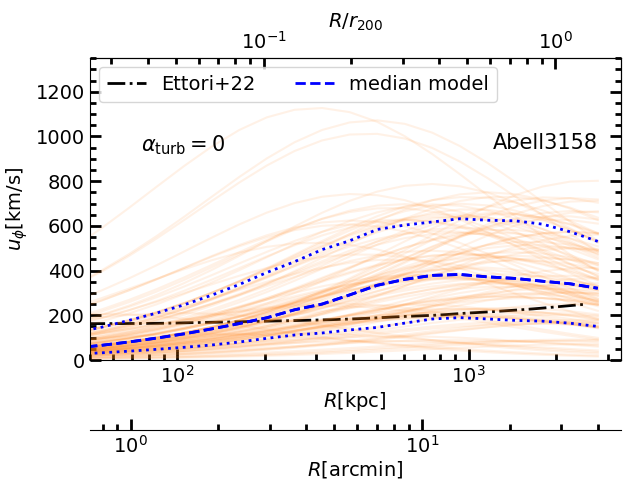}
   \includegraphics[width=0.33\textwidth]{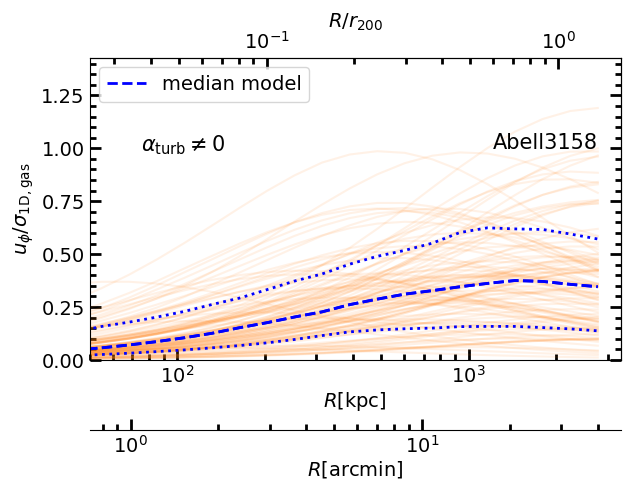}
   \includegraphics[width=0.33\textwidth]{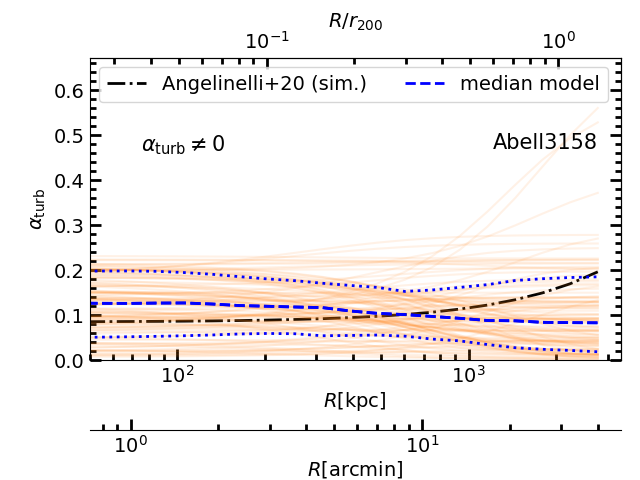}
} \includegraphics[width=0.49\textwidth]{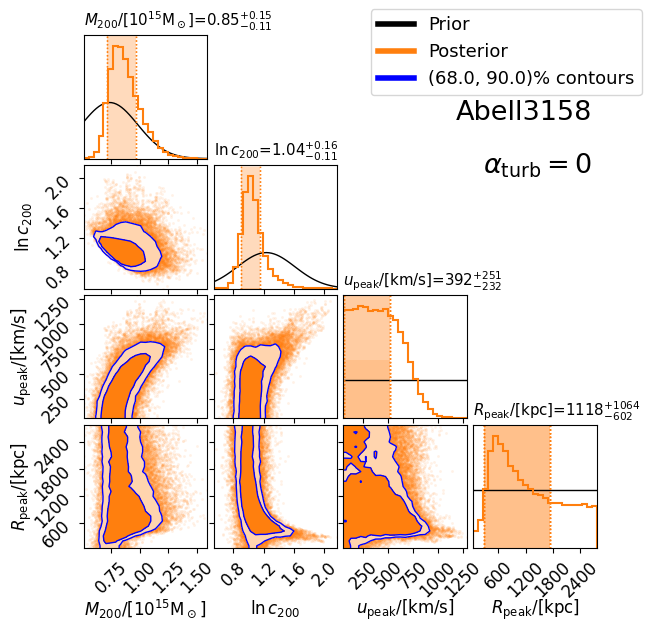}   
    \caption{Same as \ref{fig.A85}, but for A3158.}
    \label{fig.A3158}
\end{figure*}

\begin{figure*}
   \centering
   \hbox{
   \includegraphics[width=0.33\textwidth]{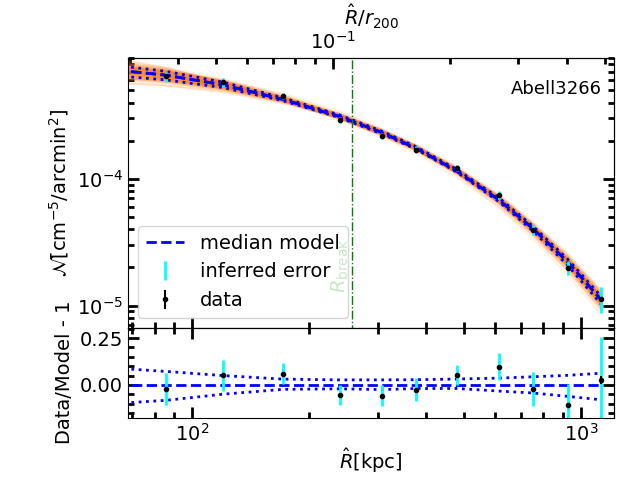}
   \includegraphics[width=0.33\textwidth]{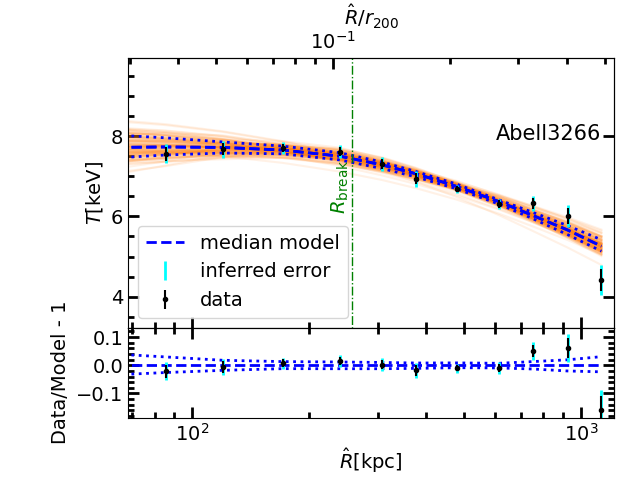}
   \includegraphics[width=0.33\textwidth]{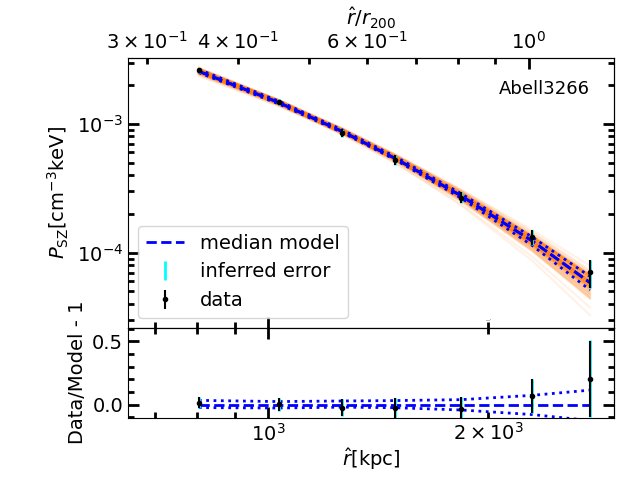}
   }
   \hbox{
   \includegraphics[width=0.33\textwidth]{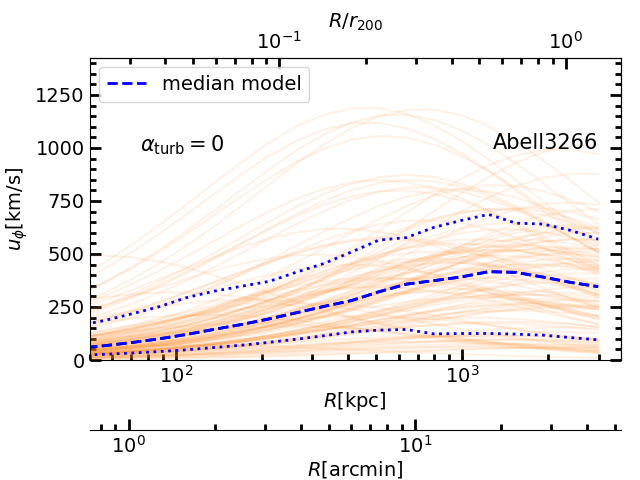}
   \includegraphics[width=0.33\textwidth]{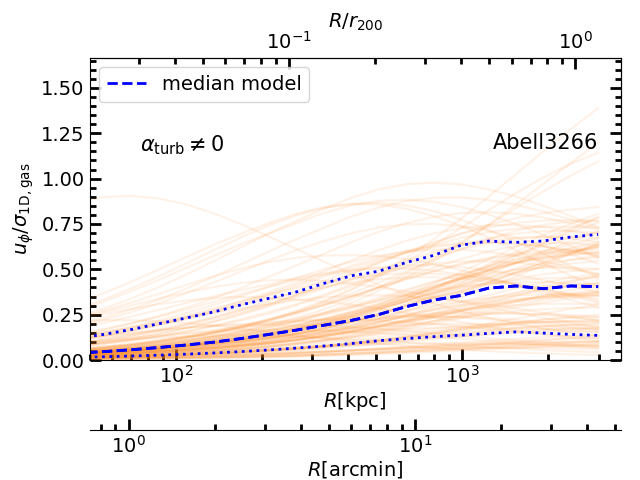}
   \includegraphics[width=0.33\textwidth]{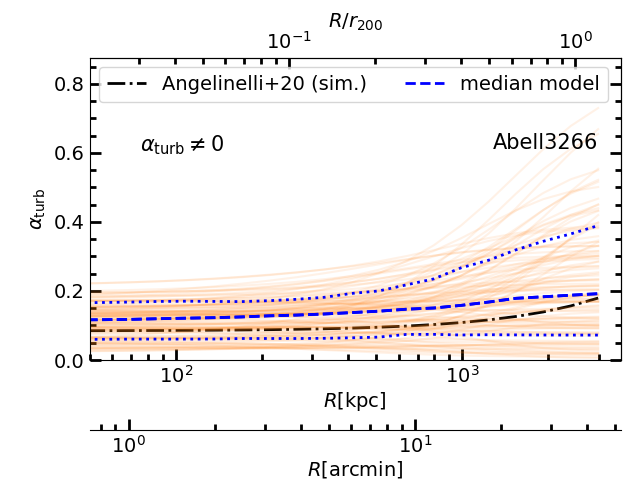}
} \includegraphics[width=0.49\textwidth]{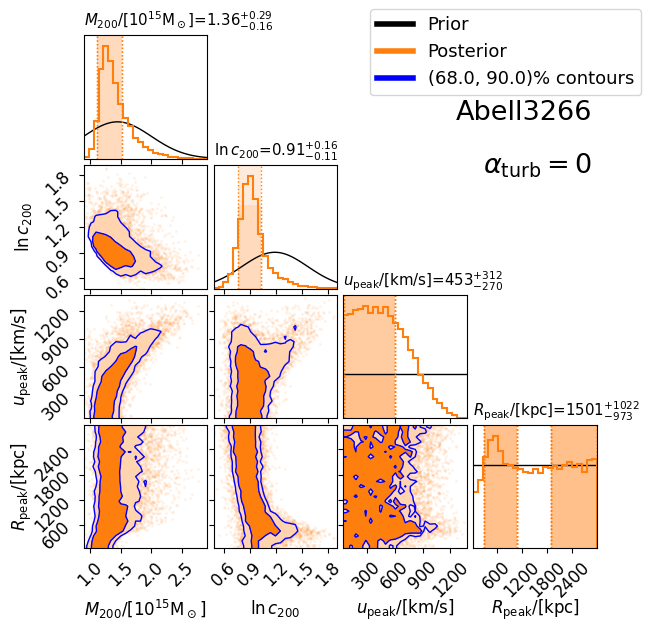}   
    \caption{Same as \ref{fig.A85}, but for A3266.}
    \label{fig.A3266}
\end{figure*}

\begin{figure*}
   \centering
   \hbox{
   \includegraphics[width=0.33\textwidth]{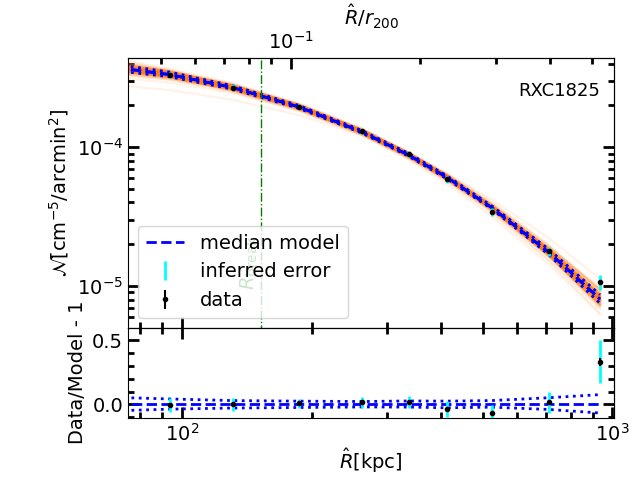}
   \includegraphics[width=0.33\textwidth]{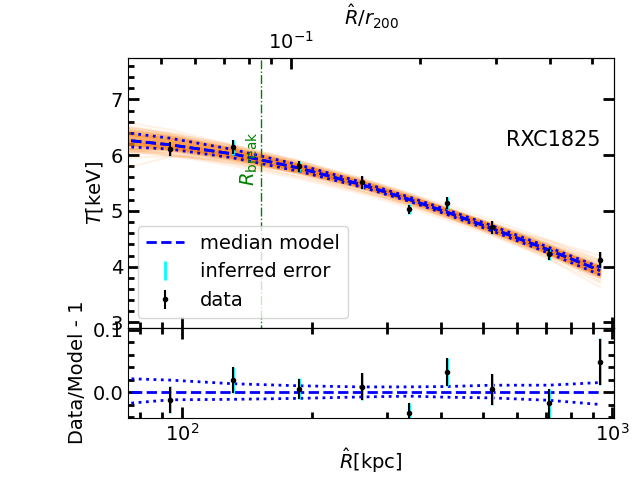}
   \includegraphics[width=0.33\textwidth]{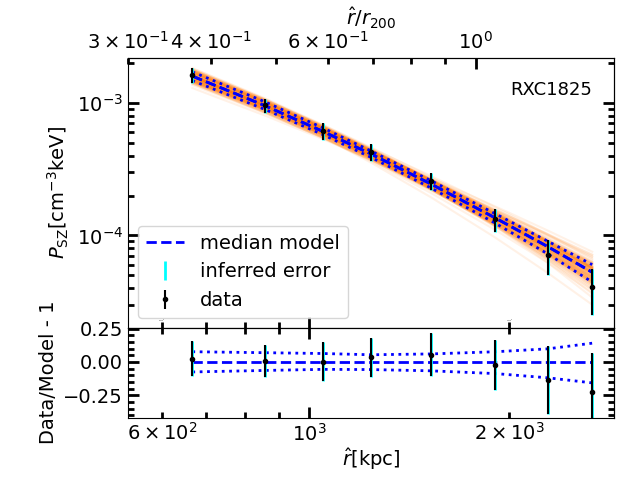}
   }
   \hbox{
   \includegraphics[width=0.33\textwidth]{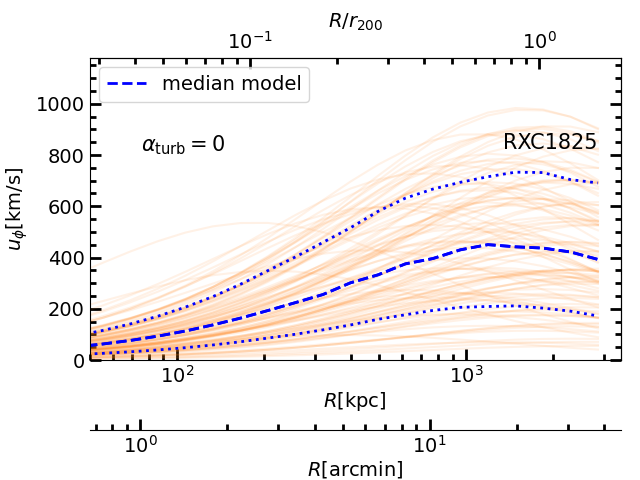}
   \includegraphics[width=0.33\textwidth]{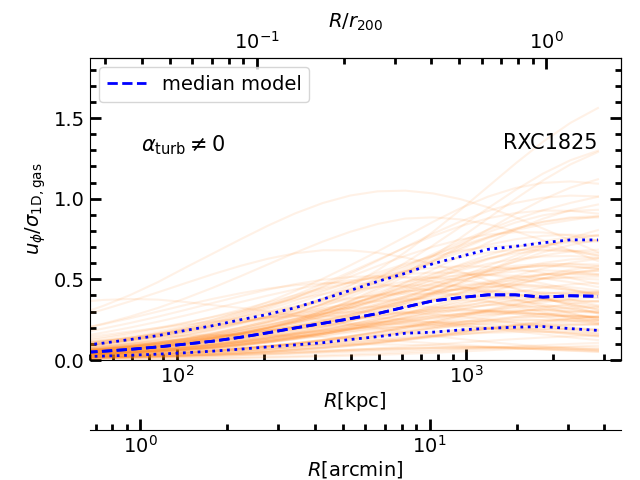}
   \includegraphics[width=0.33\textwidth]{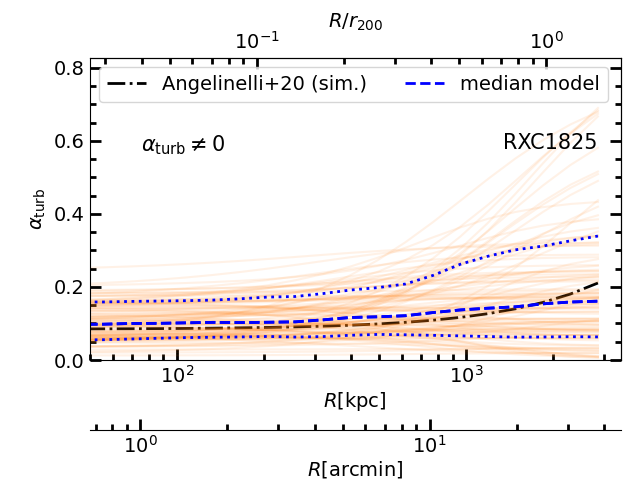}
} \includegraphics[width=0.49\textwidth]{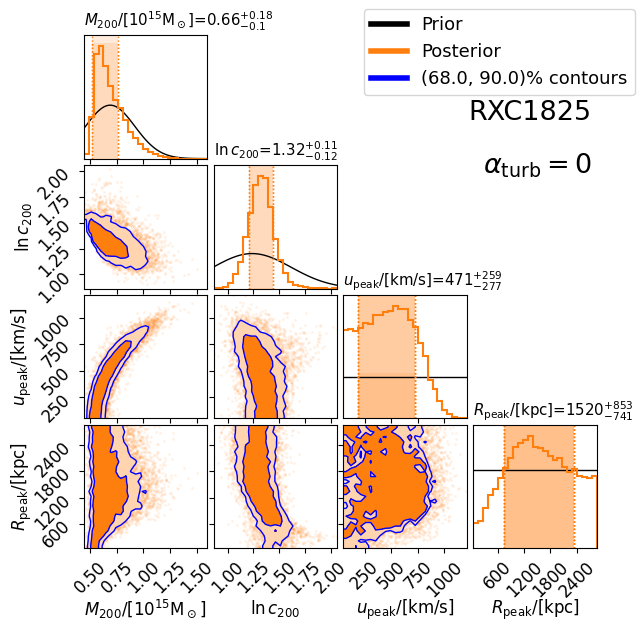}   
    \caption{Same as \ref{fig.A85}, but for RXC1825.}
    \label{fig.RXC1825}
\end{figure*}

\begin{figure*}
   \centering
   \hbox{
   \includegraphics[width=0.33\textwidth]{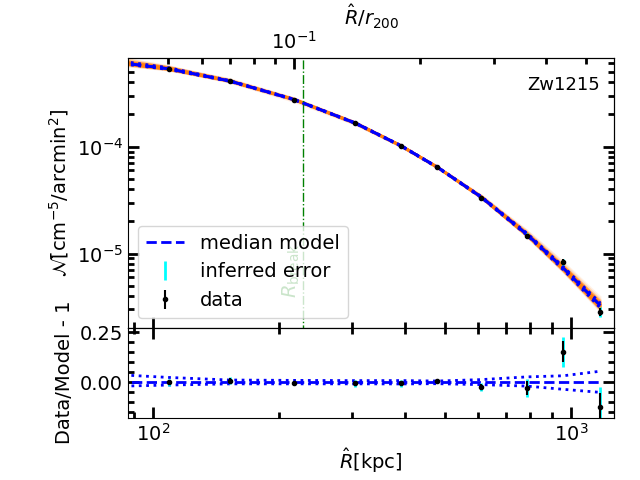}
   \includegraphics[width=0.33\textwidth]{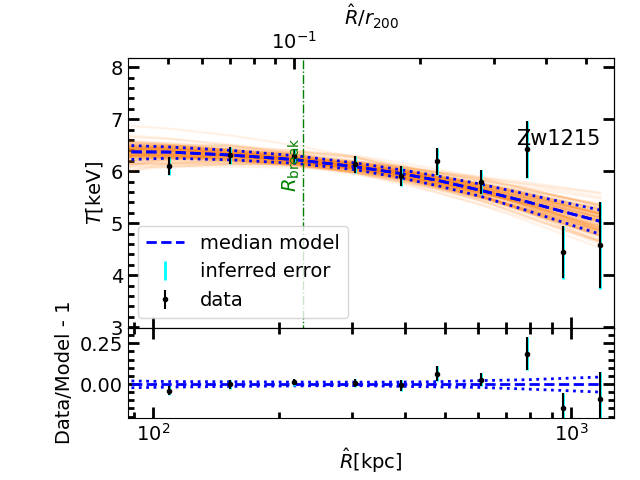}
   \includegraphics[width=0.33\textwidth]{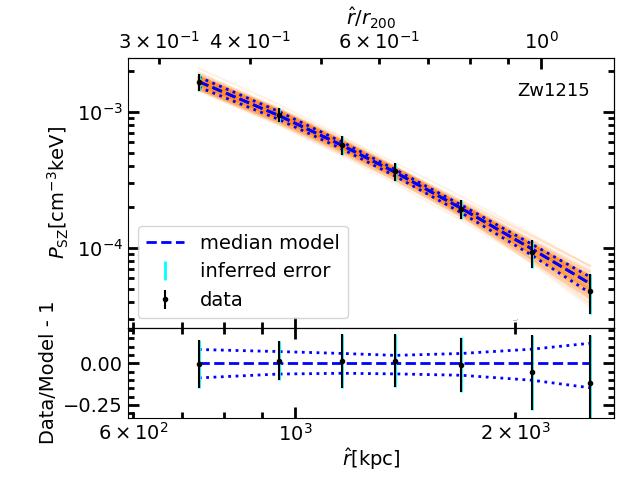}
   }
   \hbox{
   \includegraphics[width=0.33\textwidth]{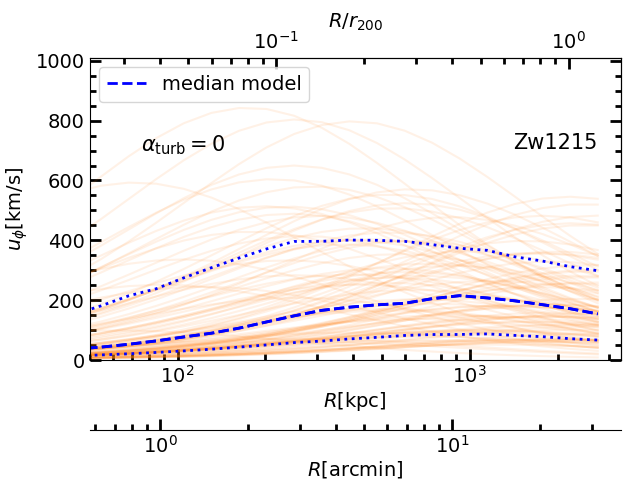}
   \includegraphics[width=0.33\textwidth]{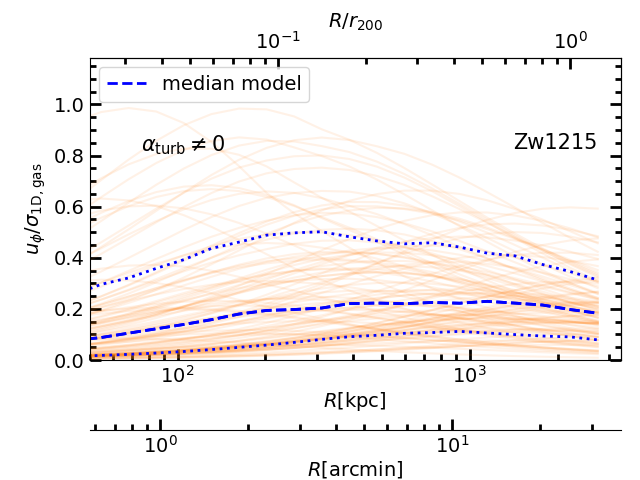}
   \includegraphics[width=0.33\textwidth]{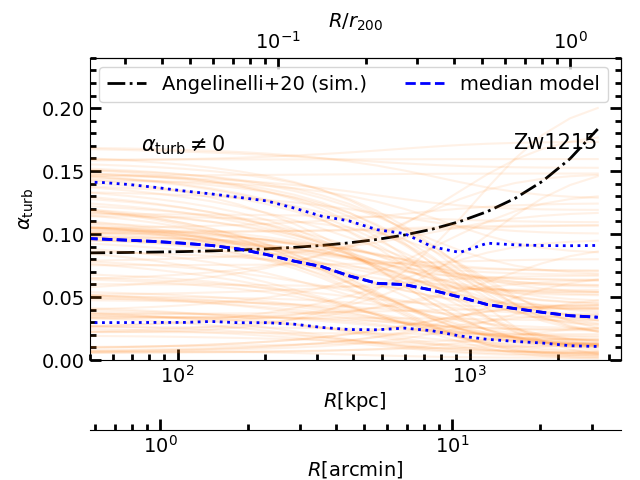}
} \includegraphics[width=0.49\textwidth]{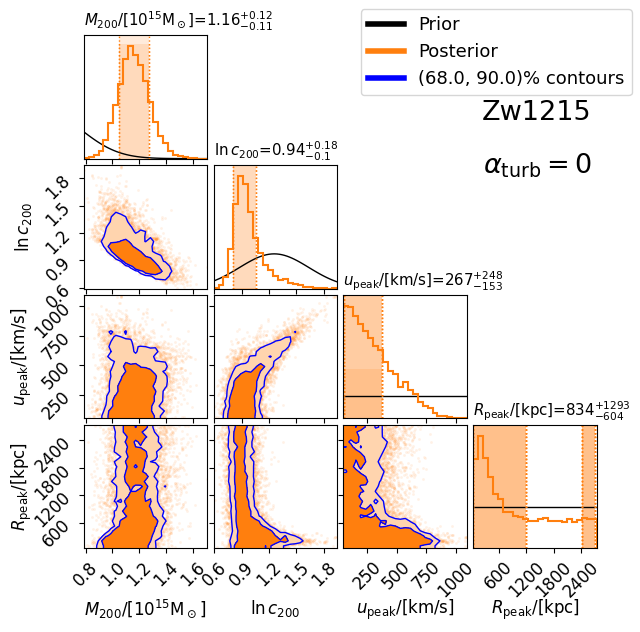}   
    \caption{Same as \ref{fig.A85}, but for Zw1215.}
    \label{fig.Zw1215}
\end{figure*}

\end{appendix}

\end{document}